\documentclass[aps,prd,reprint,superscriptaddress,nofootinbib,amsfonts,amssymb,amsmath]{revtex4-1}
 
\usepackage{dsfont}
\usepackage{graphicx}
\usepackage[colorlinks]{hyperref}
\usepackage{xcolor}
\usepackage{bbm}
\usepackage{subfigure}

\hypersetup{
colorlinks=true,
citecolor=blue,
citebordercolor=red,
linktoc=all,
linkcolor=blue
}

\allowdisplaybreaks

\graphicspath{{./figs/}}

\usepackage{color}

\definecolor{bjcol}{rgb}{1,.44,0.13}

\def\CF{{\mathcal F}}

\def\eq#1{(\ref{#1})}

\def\0#1#2{\frac{#1}{#2}}
\def\eq#1{(\ref{#1})}

\newcommand{\EH}{\text{\EH}}

\begin{document}

\title{Curvature dependence of quantum gravity with scalars}

%
\author{Benjamin~B\"urger}
\affiliation{Institut f\"ur Theoretische Physik, Justus-Liebig-Universit\"at Gie\ss en,
35392 Gie\ss en, Germany}
\author{Jan~M.~Pawlowski}
\affiliation{Institut f\"ur Theoretische Physik, Universit\"at Heidelberg,
Philosophenweg 16, 69120 Heidelberg, Germany}
\affiliation{ExtreMe Matter Institute EMMI, GSI Helmholtzzentrum f\"ur
Schwerionenforschung mbH, Planckstr.\ 1, 64291 Darmstadt, Germany}
\author{Manuel~Reichert}
\affiliation{CP$^3$-Origins, University of Southern Denmark, Campusvej 55, 5230 Odense M, Denmark}
\author{Bernd-Jochen~Schaefer}
\affiliation{Institut f\"ur Theoretische Physik, Justus-Liebig-Universit\"at Gie\ss en,
35392 Gie\ss en, Germany}
%

\begin{abstract}
  We compute curvature-dependent graviton correlation functions and couplings
  as well as the full curvature potential $f(R)$ in asymptotically safe quantum gravity coupled to scalars.
  The setup is based on a systematic vertex expansion about metric backgrounds with constant curvatures 
  initiated in \cite{Christiansen:2017bsy} for positive curvatures. We extend these results to negative 
  curvature and investigate the influence of minimally coupled scalars.
  The quantum equation of motion has two solutions for all accessible numbers of scalar fields.
  We observe that the solution at negative curvature is a minimum, while the solution at positive curvature is a maximum.
  We find indications that the solution to the equation of motions for scalar-gravity systems is at large positive curvature, 
  for which the system might be stable for all scalar flavours.
  \\
[.3cm]
{\footnotesize  \it Preprint: CP$^3$-Origins-2019-44 DNRF90}
\end{abstract}

\maketitle

\section{Introduction}
Asymptotically safe gravity \cite{Weinberg:1980gg} is a highly interesting candidate theory for 
quantum gravity. It relies on an interacting ultraviolet (UV) fixed point of the renormalisation 
group (RG) flow, the Reuter fixed point \cite{Reuter:1996cp}. It renders the high-energy behaviour of gravity finite. 
For reviews see \cite{Niedermaier:2006wt,Litim:2011cp,
	Reuter:2012id,Bonanno:2017pkg,Eichhorn:2017egq,Percacci:2017fkn,Wetterich:2019qzx}.

The physics of asymptotically safe gravity is encoded in the full, diffeomorphism invariant 
quantum effective action of the theory, $\Gamma[g]$. Its $n$-point functions, evaluated on the quantum equations of motion (EoM), 
describe graviton correlation functions, from which general observables can be constructed. Accordingly, the knowledge of the quantum EoM is chiefly important for computing observables. However, the quantum EoM is also very interesting by itself as it provides the 
non-trivial dynamical background metric, pivotal for the physics of the early universe, inflation and a resolution of the cosmological constant problem. 

In most approaches to quantum gravity the diffeomorphism invariant action $\Gamma[g]$ can only be computed 
within a split of the full metric, typically a linear split $g_{\mu\nu}= \bar g_{\mu\nu}+ h_{\mu\nu}$. Here, 
$\bar g_{\mu\nu}$ is a generic background and $h_{\mu\nu}$ are quantum fluctuations about this background. 
This leads us to an effective action $\Gamma[\bar g, h]$ with the diffeomorphism invariant action $\Gamma[g]=\Gamma[g,0]$. 
In most approaches, the split is inevitable and allows to compute the propagation and scattering of fluctuations $h$, hence the quantum 
gravity effects, in terms of background covariant momentum modes. Accordingly, this choice is part of a gauge fixing, 
and the gauge and background independence of the diffeomorphism invariant effective action encode physical diffeomorphism invariance and are hence chiefly important for quantum gravity approach. Moreover, the diffeomorphism invariant effective action can only be computed from the gauge-fixed correlation functions of the fluctuation field. This is simple to see as the latter carries the quantum fluctuations. 
While also the correlation functions of the fluctuation field carry the full quantum dynamics of the theory, it is rather difficult to 
directly construct observables from it. This leaves us with the task to compute $\Gamma[g]$ from the background-dependent correlation functions of $h$. 

In the present work, this is done with the functional renormalisation group approach \cite{Wetterich:1992yh}, leading to exact and closed one-loop 
relations between full correlation functions. While the background metric $\bar g_{\mu\nu}$ can be kept generic in these relations, it is technically challenging and limits the size of the truncation. In almost all computations that disentangle background and fluctuation field,
a flat Euclidean background was chosen, see, e.g., \cite{Christiansen:2012rx,Christiansen:2014raa,Christiansen:2015rva,Christiansen:2016sjn,Denz:2016qks,Eichhorn:2018ydy,Eichhorn:2018nda}, which is technically highly advantageous, and the computations resemble 
standard non-perturbative ones. We emphasise that the correlation functions in the flat background can be used to construct the diffeomorphism invariant action. However, from the viewpoint of convergence of an expansion in powers of $h$, an expansion about or close to the minimal solution to the EoM is much wanted, see \cite{Christiansen:2017bsy,Knorr:2017fus,Knorr:2017mhu}. Hence, getting access to $\Gamma[g]$ and the solutions to the EoM serves a twofold purpose. 
It allows us to directly access the interesting physics questions mentioned above as well as answering the question of how close the flat metric is to the full dynamical metric that solves the EoM. 

The computation of full, background-dependent correlation functions of the fluctuation field, and the computation of the diffeomorphism invariant action from the correlation functions of the fluctuation field has been initiated in \cite{Christiansen:2017bsy}. There the $f(R)$-potential for backgrounds with positive constant curvature has been computed, for respective results within the background field approximation see, e.g., \cite{Codello:2007bd,Machado:2007ea,Codello:2008vh,Benedetti:2012dx,Dietz:2012ic,Falls:2013bv,Benedetti:2013jk,Dietz:2013sba,Demmel:2014sga,Falls:2014tra,Demmel:2015oqa,Ohta:2015efa,Ohta:2015fcu,Falls:2016wsa,Falls:2016msz,Gonzalez-Martin:2017gza,deBrito:2018jxt,Alkofer:2018baq,Falls:2018ylp}. In the present work we extend the computation to negatively curved backgrounds, which allows us to study global minima. Moreover, we also study the impact of matter fields to the system by adding minimally coupled scalar fields.
Scalar-gravity systems have attracted a lot of attention
\cite{Dona:2013qba,Meibohm:2015twa,Dona:2015tnf,Eichhorn:2018akn,Alkofer:2018fxj,
  Henz:2013oxa,Percacci:2015wwa,Labus:2015ska,Oda:2015sma,Henz:2016aoh,Wetterich:2016uxm,Biemans:2017zca,
  Hamada:2017rvn,Becker:2017tcx,Eichhorn:2017sok,Eichhorn:2017als,Alkofer:2018fxj,Pawlowski:2018ixd,Knorr:2019atm,Wetterich:2019rsn},
in particular, since these systems seemingly have a divergence in the Newton coupling \cite{Dona:2013qba,Meibohm:2015twa,Dona:2015tnf,Eichhorn:2018akn}.
As discussed in \cite{Meibohm:2015twa,Christiansen:2017cxa}, this is expected to be an artefact of the truncation.
We discuss this issue for the first time for curvature dependent couplings.
Indeed, we find indications that the best expansion point for scalar-gravity systems is at large positive curvature,
which might solve the stability issues encountered in previous works.

This paper is structured as follows.
In \autoref{sec:gen-framework} be briefly explain our ansatz for the graviton $n$-point functions and the use of the functional renormalisation group (FRG). 
In \autoref{sec:curved-background} we recap the approximation of the momentum space on curved backgrounds, initiated in \cite{Christiansen:2017bsy},
and describe the evaluation of traces of the Laplacian on positive and negative background curvatures.
The fixed-point solutions and their dependence on the number of scalars are displayed in \autoref{sec:AS}.
In \autoref{sec:EoMs} we discuss the background and the quantum EoM, their solutions and their asymptotic behaviour.
We summarise our findings in \autoref{sec:summary}. 

\section{General Framework}
\label{sec:gen-framework}
We start from the gauged-fixed Einstein-Hilbert action with $N_s$ minimally coupled scalar fields,  
\begin{align}
	\label{eq:EHaction}
	S_{\text{EH}}={}&\frac{1}{16\pi G_\text{N}}\int \text{d}^4 x \sqrt{g}\left(2
	\Lambda-R\right)+S_{\text{gf}}+S_\text{gh} \notag \\[1ex]
	&+\sum_{i=1}^{N_\text{s}}\frac{1}{2}\int \text{d}^4x\sqrt{g}\,\nabla_\mu\varphi_i\nabla^\mu\varphi_i \,.
\end{align}
The action in \eqref{eq:EHaction} is expanded about a maximally symmetric background metric with background curvature $\bar R$. The gauge fixing is given by 
\begin{align}\nonumber 
	S_\text{gf} =\01{2\alpha}\int\mathrm d^4x\,
	\sqrt{\bar g}\,\bar g^{\mu\nu}F_\mu F_\nu\,,\\[1ex] 
	S_\text{gh}=\int\mathrm d^4 x \,\sqrt{\bar{g}}\,  
	\bar g^{\mu \mu'} \bar g^{\nu \nu'}\bar{c}_{\mu'}
	{\cal M}_{\mu\nu} c_{\nu'} \,.
\end{align} 
with the Faddeev-Popov operator ${\cal M}_{\mu\nu}(\bar g, h)$ of the gauge
fixing $F_\mu(\bar g, h)$. We employ a linear, de-Donder type gauge fixing,
\begin{align} \nonumber F_\mu ={}& \bar \nabla^\nu
	h_{\mu\nu} - \0{1+\beta}4 \bar \nabla_\mu {h^\nu}_\nu \,, \\[1ex] 
	{\cal M}_{\mu\nu} ={}&
	\bar\nabla^\rho\left(g_{\mu\nu}\nabla_\rho+g_{\rho\nu}\nabla_\mu
	\right)-\bar\nabla_\mu\nabla_\nu\,.  
	\label{eq:gravitygaugefixing} 
\end{align}
We work with $\beta=0$ and the Landau-deWitt gauge limit $\alpha\to0$, which is a fixed point
of the RG flow \cite{Litim:1998qi}, for more details see e.g.~\cite{Christiansen:2017cxa, Christiansen:2017bsy}. 
We use a linear split of the full metric $g_{\mu\nu}$ abound a background $\bar g_{\mu\nu}$ with constant curvature and the fluctuation $h_{\mu\nu}$,  
\begin{align}
	g_{\mu\nu}=\bar{g}_{\mu\nu}+\sqrt{Z_h G_\text{N}}\,h_{\mu\nu}\,.
\end{align}
The fluctuation field is rescaled with a wave-function renormalisation $Z_h$ and the Newton coupling $G_\text{N}$. The latter gives the field the mass dimension one. Note that our setup also allows for expansions about more general backgrounds. In the present work we restrict ourselves to backgrounds with constant curvature for simplicity.

The main objective of  this work is the computation of the gauge-fixed correlation functions of the fluctuations fields, 
\begin{align}
	\frac{\delta^n \Gamma_k[\bar g,\Phi]}{\delta \Phi_1\cdots \delta\Phi_n}\equiv\Gamma_k^{(\Phi_1...\Phi_n)}\,,
\end{align}
where $\Phi=\{h_{\mu\nu},c_{\mu},\bar c_\mu, \varphi\}$. These correlation functions can be expanded in respective general tensor bases which can be significantly reduced by Slavnov-Taylor or Ward identities for the tensors. This reduces the general tensor basis to one which can be mapped to the tensor basis derived from diffeomorphism invariant terms. In the current work we only consider the leading classical tensor structure for $n$-point functions, the Einstein-Hilbert tensor structures derived by $n$ metric derivatives from \eqref{eq:EHaction}.
At each order we replace Newton's coupling and the cosmological constant by their corresponding level $n$ couplings $G_n$ and $\Lambda_n$  \cite{Christiansen:2014raa,Christiansen:2015rva,Christiansen:2017bsy}.
Their respective flow equations are obtained by differentiating the Wetterich equation \cite{Wetterich:1992yh,Ellwanger:1993mw,Morris:1993qb} $n$
times with respect to the fluctuation field.  The Wetterich equation with the given field content
is given by
\begin{align}  \label{eq:Wetterich}
  \partial_t\Gamma_k=&\,
  \frac12 \text{Tr}\, G_{k,hh}\,\partial_t R_{k,h}
  - \text{Tr}\, G_{k,\bar{c}c}\,\partial_t R_{k,c} 
  \notag \\[1ex]
  &\,+
  \012 \text{Tr}\,G_{k,\varphi\varphi}\,\partial_t R_{k,\varphi} \,,
\end{align}
with the regularised fluctuation propagators 
\begin{align}
G_{k,\Phi_1\Phi_2}=\left[\0{1}{\Gamma_k^{(\Phi\Phi)}+R_{k}}\right]_{\Phi_1\Phi_2}\,.
\end{align}
In \eqref{eq:Wetterich}, $\partial_t$ is the derivative with respect to RG-time $t = \ln k/k_0$,
where $k_0$ is some reference scale.
We truncate and close the flow equations for the higher vertices by
setting $G \equiv G_{n \geq 3}$ and $\Lambda_4 = \Lambda_3$ with
$\Lambda_{n >4} =0$. From the corresponding $n$-point function we
concentrate in the following on the curvature dependent mass-parameter
$\mu(r)$ (from the graviton two-point function), the gravitational
coupling $g(r)$ (from the three-point function) and the
momentum-independent coupling $\lambda_3(r)$, which are dimensionless and defined by 
\begin{align}
r & = \bar{R} /k^2 \,,
&
g & = G\,k^2 \,, \notag \\[1ex]
\mu & = -2 \Lambda_2 /k^2 \,,
&
\lambda_3 & = \Lambda_3 /k^2 \,.
\end{align}
Here, $\bar R$ denotes the background curvature. 
The beta functions for these couplings are obtained by an appropriate projection procedure, 
where we concentrate on the transverse traceless part of the flow, see \cite{Christiansen:2017bsy}.
The anomalous dimensions for all fields are set to zero.

We choose the regulator $R_k$ proportional to the two-point functions at vanishing
cosmological constant and background curvature
\begin{align}
 R_k = \Gamma_k^{(2)}(\mu=0,r=0) \cdot r_k(\bar \nabla_\mu^2/k^2)\,.
\end{align}
The shape function $r_k$ is chosen to be an exponential
\begin{align}
 r_k(x) = \frac{\mathrm e ^{-x^2}}{x}\,.
\end{align}
The gravity parts of the flows for $g(r)$, $\mu(r)$ and $\lambda_3(r)$ are the same as in \cite{Christiansen:2017bsy}, 
see App.~B therein. In the present matter-gravity system with minimally coupled scalars, the flows of the graviton $n$-point functions receive contributions from the scalars. Thus all flow equations depend on the number of scalar fields $N_s$. For $r=0$, this dependence was already investigated in \cite{Meibohm:2015twa,Eichhorn:2018akn}.
Here, we extend the analysis and consider the curvature dependence of the couplings.

\section{Vertices and trace evaluation on curved backgrounds}
\label{sec:curved-background}
Propagators on a non-trivial background metric $\bar{g}$ depend on the 
Laplacian $\Delta_{\bar{g}} \equiv -\bar{\nabla}^2$ and on curvature invariants. 
For the hyperbolic or spherical background considere here, the dependence on the 
curvature invariants reduces to a dependence on the background Ricci scalar $\bar R$,
$G \equiv G(\Delta_{\bar g},\bar R)$. All Laplacians can be expressed by the
scalar Laplacian and the Ricci scalar.

The higher-order vertices $\Gamma^{(n \geq 3)}$ also depend on
covariant derivatives $\bar{\nabla}_{\mu}$.  In curved backgrounds the
Laplacian and the covariant derivative do not commute and the lack of
a common eigenbasis also determines the lack of a momentum space.  As
in \cite{Christiansen:2017bsy} we construct an approximate momentum
space.  All covariant derivatives are symmetrised, which produces
further background curvature terms
\begin{align}
  \bar{\nabla}^{\mu}\bar{\nabla^\nu}&=
  \frac{1}{2}\lbrace \bar{\nabla}^{\mu},\bar{\nabla}^{\nu}\rbrace
  +\bar{R}\text{-terms}\,.
\end{align} 
The symmetrised covariant derivatives are expressed by the spectral 
value of the Laplacian $p^2$ and an spectral angle $x$ between them
\begin{align}
  \bar{\nabla}_1 \cdot \bar{\nabla}_2 = x \sqrt{p_1^2} \sqrt{p_2^2} \,.
\end{align}
The spectral angle depends non-trivially on the spectral values of the
covariant derivative and the background curvature.
Until here, we have not performed any approximation and we have simply 
stored the background curvature dependence in the spectral angle.
Now we approximate the spectral angle with the corresponding
flat spacetime angle $x\approx\cos \theta_\text{flat}$.
The flat spacetime angles are integrated using the usual volume element 
of the spacetime, such that the approximation becomes exact
in the limit $\bar{R}\rightarrow 0$.
In \cite{Christiansen:2017bsy} it was estimated that this approximation should 
give reasonable results for $|\bar{R}/k^2| \lesssim 2$.
The external spectral values of
the three-point function are evaluated at the symmetric point
\begin{align}
  p:=|p_1|=|p_2|=|p_3| \,,
\end{align}
with the flat angle $\theta_{\text{flat}}=\frac{2\pi}{3}$ between
them. This procedure leaves us with functions that depend on the
Laplacian and the background curvature, but not anymore on the covariant derivative.
The spectrum of the Laplacian on hyperbolic or spherical backgrounds is known and the traces can
be evaluated using spectral sums (integrals) for positive (negative)
background curvature \cite{Rubin:1984tc,Camporesi:1994ga,Percacci:2017fkn}.

\subsection{Positive curvature}
For positive background curvature the spectrum of the Laplacian is discrete.
The dimensionless eigenvalues of the scalar Laplacian are given by \cite{Rubin:1984tc}, 
\begin{align}
  \omega(\ell)=\frac{\ell(\ell+3)}{12}r \,,
\end{align}
with multiplicities
\begin{align}
  m(\ell)=\frac{(2\ell+3)(\ell+2)!}{6l!} \,,
\end{align}
for positive integers $\ell>0$. A trace of a function $F$ that depends on the Laplacian 
and the background curvature is evaluated by
\begin{align}
\frac1{V}\,  \text{Tr}F(\Delta,r)\rightarrow \frac{r^2}{384\pi^2}
  \sum_{\ell=2}^{\ell_\text{max}}m(\ell)F(\omega(\ell),r) \,.
\end{align}
The factor $V=\frac{384\pi^2}{r^2}$ is the dimensionless
four-sphere volume. We cut the sum at some finite $\ell_\text{max}$
where the sum is sufficiently converged. We further start the sum at
$\ell=2$ and exclude the modes $\ell=0,1$. This is correct for the
transverse-traceless mode of the graviton but an approximation for the trace mode. Our
procedure does not allow to distinguish between these modes. The
approximation is well justified as the low modes only contribute at
large background curvature and we are interested in the small
curvature regime.

\subsection{Negative curvature}
For negative curvature spacetime is unbounded and the 
spectrum of eigenvalues is continuous. For the scalar Laplacian, the spectrum is given by \cite{Camporesi:1994ga}, 
\begin{align}
  \lambda(\sigma)=\frac{|r|}{12}\left(\sigma^2+\frac{9}{4}\right) \,,
\end{align}
with the spectral density
\begin{align}
  \rho(\sigma)=\left[\sigma^2+\frac{1}{4}\right]\sigma\tanh(\pi\sigma)\,.
\end{align}
In this case a trace over a function $F$ yields the spectral integral
\begin{align}
 \frac1{V}\, \text{Tr} F(\Delta,r)\rightarrow\frac{1}{3}\frac{r^2}{384\pi^2} 
  \int \text{d}\sigma \rho(\sigma) F(\lambda(\sigma),r) \,, 
\end{align}
where again the volume prefactor gives the correct
$r\rightarrow 0$ limit.

\subsection{Heat-kernel expansion}
The beta functions of the coupling functions are partial differential equations
in the RG scale $k$ and the background curvature $r$.
The search for fixed-point functions reduces the beta functions to 
ordinary linear differential equations. We use a heat-kernel expansion 
about the flat background to provide initial conditions to 
these differential equations.

The heat-kernel expansion for positive and negative $r$ is,
in the case of a scalar Laplacian, given by
\begin{align}
 \frac1{V} \, \text{Tr}F(\Delta,r)&\rightarrow \frac{1}{V}\int_0^\infty \!\! \text{d}s \,
                     \text{Tr}\!\left[\mathrm e^{-s \Delta}\right]\tilde{F}(s,r) \notag \\[1ex] 
                     &
  =\frac{1}{(4\pi)^2}\left(Q_2[F]+Q_1[F]\frac{r}{6}+...\right) \,,
\end{align}
where the $Q_n$ functionals for $n>0$ are given by
\begin{align}
  Q_n[F]=\frac{1}{\Gamma(n)}\int_0^\infty  \!\text{d}\lambda\,\lambda^{n-1}F(\lambda,r) \,.
\end{align}
We compute the zeroth and first order in the background curvature in order
to have a smooth continuation to the spectral sums and integrals at finite but small $r$.

\begin{figure}[t]
	\includegraphics[width=\linewidth]{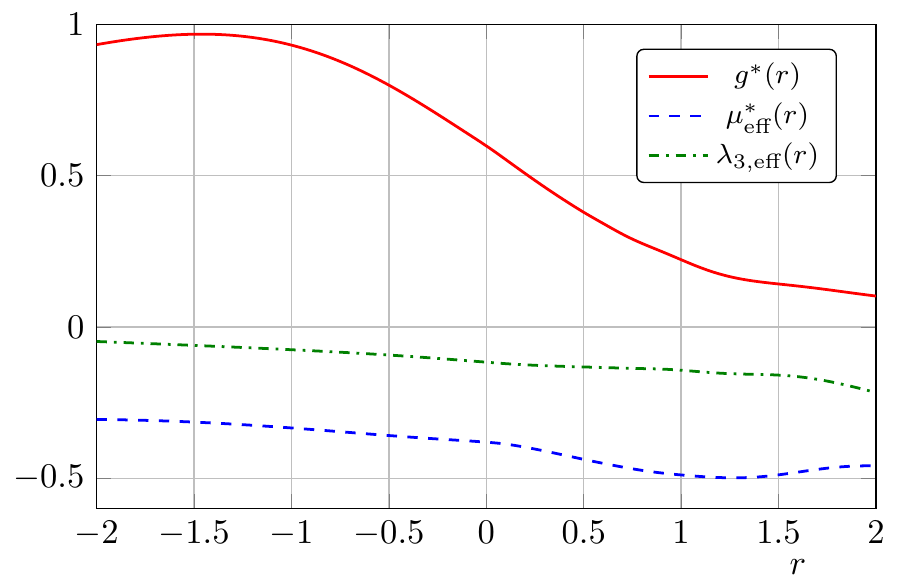}
	\caption{
	Fixed-point functions $g^*(r)$, $\mu^{*}_{\text{eff}}(r)= \mu(r)+\frac{2}{3}r $  
	and  $\lambda^{*}_{3,\text{eff}}(r)=\lambda_3(r)+\frac{1}{12}r$ for $N_s=0$.
	 }
	\label{fig:fullFPsolution}
\end{figure}

\section{Asymptotic safety}
\label{sec:AS}
\subsection{Pure gravity}
We begin the discussion with the fixed-point functions $g^*(r)$,
$\mu^*(r)$ and $\lambda^*_3(r)$ in the curvature regime $|r|\leq 2$
without additional scalars. The fixed-point functions are
defined by the roots of the corresponding beta functions
$\beta_{g_i}(r)=\partial_t g_i(r)$ as functions of $r$.  The initial
values for the beta functions are extracted from a heat-kernel
expansion up to linear order in $r$ 
\begin{align}
\begin{pmatrix} g^{*}(r) \\[1ex] \mu^{*}(r) \\[1ex] \lambda^{*}_{3}(r)
\end{pmatrix} =
\begin{pmatrix} \phantom{-}0.60 \\[1ex] -0.38 \\[1ex] -0.12
\end{pmatrix}+r
\begin{pmatrix} -0.43 \\[1ex] -0.71\\[1ex] -0.13
\end{pmatrix}
+\mathcal{O}(r^2)\,.
\end{align}
In the following we work with the effective couplings
\begin{align}
\mu_\text{eff}(r) & = \mu(r)+\frac{2}{3}r \,,
&
\lambda_{3,\text{eff}}(r) & = \lambda_3(r)+\frac{1}{12}r \,,
\end{align}
which are defined such that they include the explicit $r$ dependence
in the corresponding graviton two-/three-point function.

In \autoref{fig:fullFPsolution} we display the fixed-point functions in 
terms of these effective couplings.  Interestingly, the effective
couplings are almost curvature independent, except for the Newton coupling. 
This means that the explicit curvature dependence of the
$n$-point functions is almost exactly counterbalanced by the implicit
curvature dependence of the couplings. 
This was already shown in \cite{Christiansen:2017bsy} for positive curvature. 
Here we extend this non-trivial result to negative curvature.

We emphasise that this non-trivial approximate curvature-independence supports the 
expansion scheme about flat backgrounds in pure gravity. This implies the self-consistency of the flat-space angular approximation used here. Both properties are highly welcome and 
provide non-trivial reliability checks for existing fluctuation results.   

\begin{figure*}[t]
	\includegraphics[width=\linewidth]{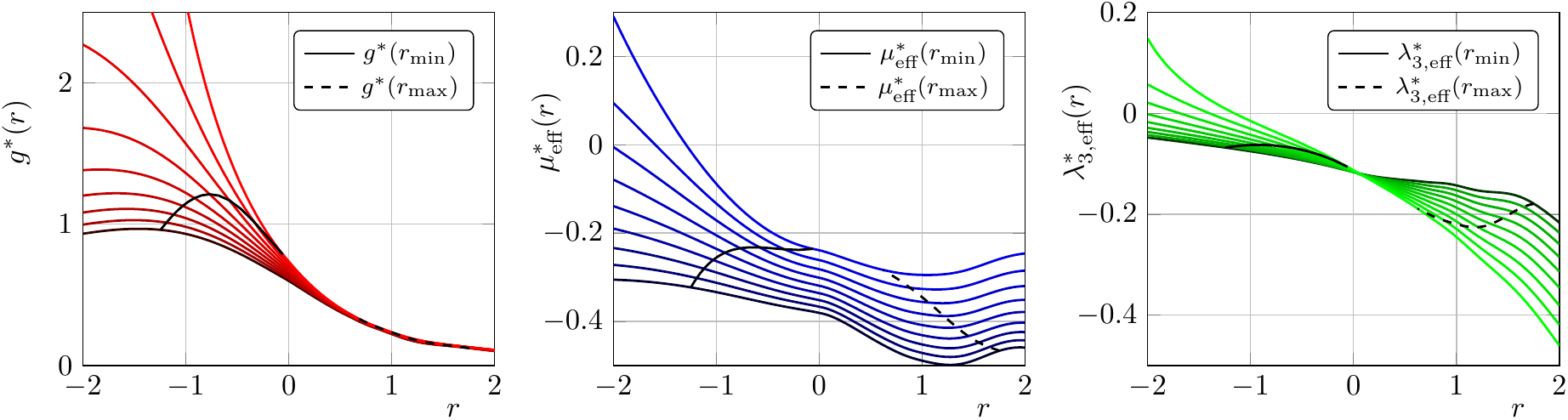}
	\caption{
	Fixed-point functions $g^{*}(r)$ (left), $\mu_\text{eff}^{*}(r)$ (middle)
	and $\lambda_{3,\text{eff}}^{*}(r)$ (right) for $N_\text{s} \in \{0,\,2,\dots,16\}$. 
	The darkest lines correspond to $N_\text{s}=0$ and lines become brighter with increasing $N_s$.
	The solid (dashed) line marks the solution to the quantum EoM, which corresponds to a minimum (maximum). 
	}
	\label{fig:scalar}
\end{figure*}

\subsection{Matter dependence}
Let us now discuss the influence of $N_\text{s}$ minimally coupled scalar
fields. This allows us to study whether the 
independence of the fluctuation correlation functions on the background 
carries over to gravity-matter systems. Moreover, gravity-scalar systems are known 
to lose the asymptotically safe fixed point in the given approximation for 
a sufficiently large number of scalars $N_s\approx 10 - 10^2$ in an expansion about the flat background, 
\cite{Dona:2013qba,Meibohm:2015twa,Dona:2015tnf,Eichhorn:2018akn,Alkofer:2018fxj}. Note in this context, that 
a heat kernel expansion in powers of the curvature is precisely the expansion about the flat background. 
However, this breakdown happens for $N_s$, 
which are already beyond hard reliability bounds of the approximations used in these works, see \cite{Meibohm:2015twa,Christiansen:2017cxa}. 
Moreover, in \cite{Christiansen:2017cxa} it has been shown that in the current approximation one should be able to map the 
theory to a pure gravity setup, hinting at further deficiencies of the FRG-treatment in the given approximation. Accordingly, it is also highly interesting to see whether the non-trivial background at least partially takes care of these deficiencies. 

To begin with, our results for the flat background $r=0$ are compatible with the 
results from \cite{Meibohm:2015twa}. There are small differences due 
to the missing anomalous dimensions, the different gauge and regulator shape function.
In the present approximation, we find that the fixed 
point at $r=0$ is vanishing at $N_\text{s}\approx 38$, corroborating earlier findings in \cite{Dona:2013qba, Meibohm:2015twa, Dona:2015tnf, Eichhorn:2018akn, Alkofer:2018fxj}. 
The full curvature dependent fixed-point functions are displayed in 
\autoref{fig:scalar}. Naively, we find an even stricter bound on the number
of scalar fields since $g^*(r)$ diverges at finite negative curvature for
$N_\text{s}>16$. 
This smaller bound is easily explained by the fact that the Newton coupling rises towards negative curvature and technically the disappearance of the 
asymptotically safe fixed point in gravity-scalar systems is related to the divergence of the Newton coupling. However, the latter 
also leads to the breakdown of the current approximation as does the limit of sufficiently large curvature $|r|$.

The divergent behaviour of $g^*(r)$ for $r<0$ is possibly triggered by
neglecting the anomalous dimension in our approximation. We have
chosen a regulator $R_k$ proportional to the two-point function and
hence proportional to the wave-function renormalisation. In the UV
the regulator scales as 
\begin{align}
	\lim\limits_{k\rightarrow \infty}R_{k,h}\sim Z_{k,h} k^2 \,,
\end{align}
which should tend to infinity. However, if the
anomalous dimension $\eta_h=-\partial_t \ln Z_{k,h}$ exceeds the
value two this is not the case anymore as the wave-function
renormalisation scales as $Z_{k,h}\sim k^{-\eta_h}$, which leads to a
vanishing regulator for $k\rightarrow \infty$. Consequently, we interpret
the divergence in $g^*(r)$ as a breakdown of the truncation and not 
as a physical bound on the compatibility of asymptotically safe gravity
with scalar fields, following the argument in \cite{Meibohm:2015twa}.  

In turn, the effects of scalar fields on $g^*(r)$ in the
positive curvature regime are small. Accordingly, as in the pure gravity the effective graviton mass
parameter $\mu^*_\text{eff}(r)$ remains almost curvature independent
and is only shifted by a constant when scalars are included. For
$\lambda^*_{3,\text{eff}}(r)$ the fixed-point value at $r=0$ is  
almost constant while for positive curvature the fixed-point function
decreases while for negative curvature it increases when additional
scalar fields are included.

\section{Background effective action and quantum equation of motion}
\label{sec:EoMs}
With the curvature-dependent fluctuation correlation functions, we can compute the fixed point diffeomorphism invariant background 
effective action $\Gamma^*[g]$. For constant curvatures the background effective action $\Gamma_k[g]=\Gamma_k[g,h=0]$ is given by 
\begin{align} \label{eq:Gamma-f}
\Gamma[{g}]  = \int \mathrm d^4x \sqrt{{g}}\, k^4 
 \tilde f(r) = V \tilde f(r) \,,
\end{align}
where $V$ is the spacetime volume and $f(R)= k^4 \tilde f(r)$. From now on, we only discuss the dimensionless function $\tilde f(r)$. We drop the tilde for being coherent with the notation in the literature. At vanishing cutoff scale, \eq{eq:Gamma-f} directly comprises the physics information of asymptotically safe quantum gravity in terms of diffeomorphism-covariant correlation functions $\Gamma^{(n)}_{k=0}$. As has been discussed in detail in \cite{Christiansen:2017bsy,Lippoldt:2018wvi}, for $k \neq 0$ there are two EoMs to be considered: 
that of the background metric 
\begin{align} 
\frac{\delta \Gamma_k}{\delta  \bar g_{\mu\nu}}\bigg|_{\bar{g}=\bar{g}_{\overline{\text{EoM}}}, h=0}=&\,0\,,
\label{eq:EoMbarg}
\end{align}
and that of the fluctuation field
\begin{align}
\frac{\delta \Gamma_k}{\delta h_{\mu\nu}}\bigg|_{\bar{g}=\bar{g}_\text{EoM}, h=0}=&\,0\,. 
\label{eq:EoMh}
\end{align}
Their respective solutions $\bar{g}_{\overline{\text{EoM}}}$  and $\bar{g}_{{\text{EoM}}}$ agree for $R_k=0$ due to the underlying diffeomorphism invariance, but they differ for $R_k\neq 0$. The latter property reflects the fact that the regularisation procedure in the functional RG breaks diffeomorphism invariance despite the persistence of diffeomorphism invariance of the background effective action: only the Slavnov-Taylor identities without the cutoff modification elevate the auxiliary background diffeomorphism invariance to physical diffeomorphism invariance carried by the symmetry properties of the fluctuation field. This leads to the counter-intuitive fact that for $k\to\infty$ it is arguably the EoM for the fluctuation field carries the physics information. In turn, the background EoM is regulator-dependent due to its dependence on the background metric, while this dependence is sub-dominant for the fluctuation EoM. 

\subsection{Background equation of motion}
For constant curvatures, as in \eqref{eq:Gamma-f}, the background EoM is given by
\begin{align}
 \label{eq:explicit-beom}
 \Gamma_k^{(\bar g_\text{tr})}[{g},0] \sim  r f'(r) - 2 f(r) = 0 \,.
\end{align} 
The flow equation for $f(r)$ is given by
\begin{align} 
\partial_t f(r) + 4 f(r) - 2 r f'(r) 
=&\,  \mathcal{F}\left(r, \mu(r),N_s\right)\,, 
\label{eq:Flowf}
\end{align}
where we have denoted the right-hand side of the Wetterich equation with $\mathcal{F}$, including the volume factor from the left-hand side.
Intriguingly, at the fixed point, $\partial_t f =0$, the left-hand side is proportional to the background EoM \eqref{eq:explicit-beom}.
Thus a root in $\mathcal{F}$ corresponds to a solution of the background EoM.
The function $\mathcal{F}$ depends on $\mu(r)$ and $N_s$ but not on the fixed-point functions of the three-point vertex, $g(r)$ and $\lambda_3(r)$.
In the computation of $\mathcal{F}$ we neglect the zero mode of the trace.
This mode develops an unphysical pole in our approximation, which would not happen if we would disentangle $\mu_\text{tt}$ and $\mu_{tr}$
or if we would use a more extended ansatz for the bare action \eqref{eq:EHaction}.
The zero mode only contributes significantly for large $r$ and thus this is a good approximation.

\begin{figure}[t]
	\includegraphics[width=\linewidth]{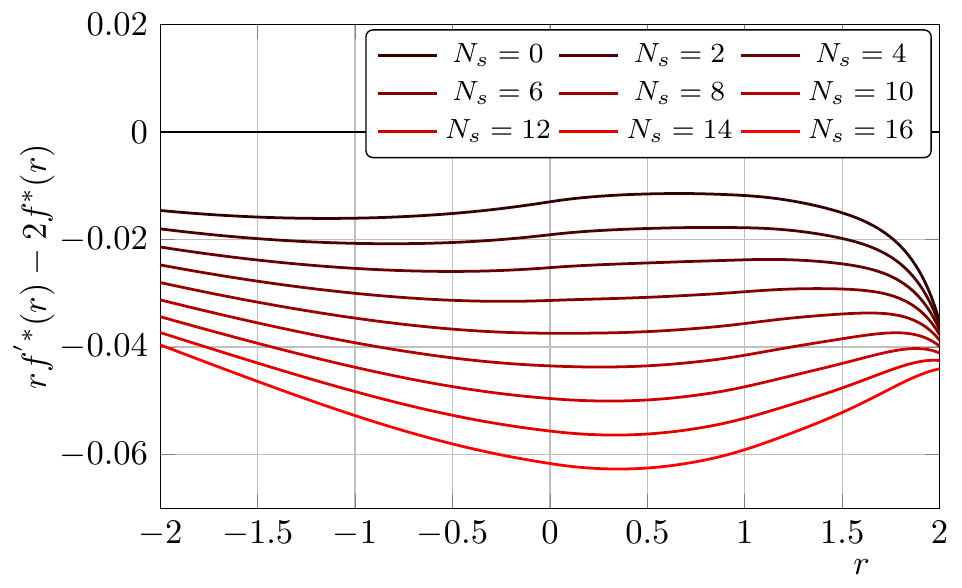}
	\caption{Background EoM for 
	different number of scalar fields.
	}
	\label{fig:bEoM}
\end{figure}

We display the function $ r f'(r) - 2 f(r)$ at the fixed point  in \autoref{fig:bEoM}.
A zero of this function corresponds to a solution of the background EoM, see \eqref{eq:explicit-beom}.
There are no solutions to the background EoM for any background curvature and any number of scalar flavours.
This is extending the results from \cite{Christiansen:2017bsy}, where the same result was found for positive curvature and without scalar fields.
We can compare this to results in the background field approximation.
There different results have been obtained depending on the choice of regulator and parameterisation.
For example in \cite{Falls:2014tra} with the linear split, only a solution at large negative curvature was found, compatible with our results.
However, in \cite{Falls:2017lst,Falls:2018ylp}, two further solutions at positive curvature were found due to a different choice of regulator.
A solution at positive curvature was also found in \cite{Demmel:2015oqa} and with the exponential parameterisation in \cite{Ohta:2015fcu}.
Importantly, we see that the existence of a solution to the EoM might crucially depend on the matter content of the theory.
This was already partially discussed in \cite{Christiansen:2017bsy,Alkofer:2018baq}.

\subsection{Quantum equation of motion}
We now turn to the quantum EoM given in \eqref{eq:EoMh}, which is the more important EoM for the reasons discussed above.
The fluctuation one-point function can be parameterised analogously to the background effective action in \eqref{eq:Gamma-f}.
Only the trace-part is non-vanishing for constant curvature metrics. We arrive at 
\begin{align}
\label{eq:Gamma-f1}
\Gamma_k^{(h_\text{tr})}[{g},0]=\int \text{d}^4 x\sqrt{g}\,  k^3 f_1(r)=\frac{V}{k} f_1(r) \,,
\end{align}
where the spatial integration can be performed because the curvature
scalar is constant. The flow of $\Gamma_k^{(h_{\text{tr}})}$ is obtained
in the same way as for the two- and three-point functions and at the
fixed point we are left with an ordinary differential equation for
$f^*_1(r)$. The initial conditions for $f_1(r)$ are again obtained by
a heat-kernel expansion. The quantum EoM reduces to
\begin{align}
  \label{eq:explicit-qeom}
  \Gamma_k^{(h_\text{tr})}[{g},0] \sim f_1(r) = 0 \,.
\end{align}
The flow equation for $f_1(r)$ is given by 
\begin{align}
\partial_t f_1(r)+ 3 f_1(r) - 2 r {f_1}'(r) =&\, \mathcal{F}_1(r, \mu(r),N_s)\,,
\label{eq:Flowf1}
\end{align}
where we have denoted the right-hand side of the Wetterich equation by $\CF_1$ again including the volume factor from the left-hand side.
Note the difference to \eqref{eq:Flowf} due to the mass dimension of the fluctuation field.

\begin{figure}[t]
	\includegraphics[width=\linewidth]{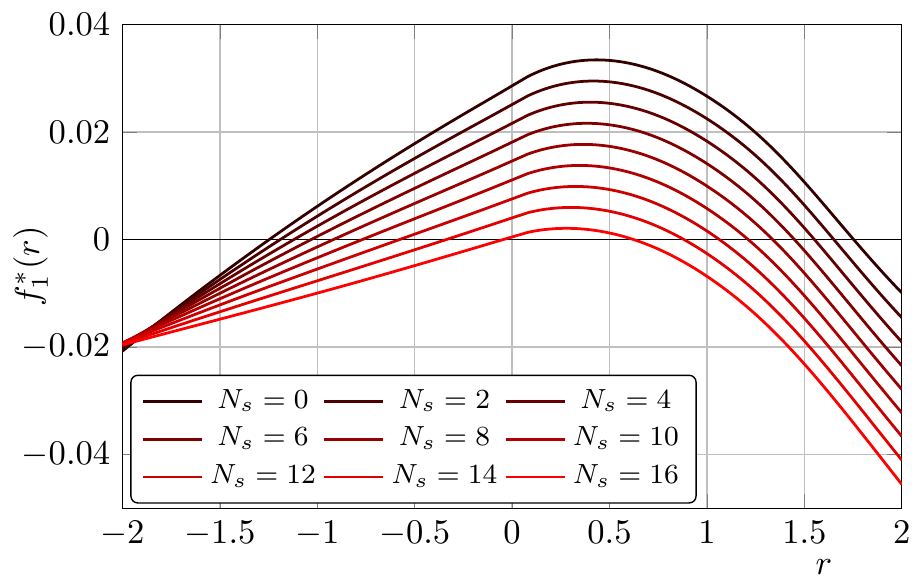}
	\caption{Quantum EoM for 
	different number of scalar fields.
	}
	\label{fig:qEoM}
\end{figure}

We display the fixed-point functions $f_1^*$ in \autoref{fig:qEoM}.
For all $N_s$, we find two solutions to the quantum EoM, one at negative curvature and one at positive curvature.
It turns out that the solution at negative curvature is a minimum, while the one as positive curvature is a maximum.
For increasing $N_s$ these solutions move towards each other, however,
just before they merge, we lose the solution of the fixed-point functions as shown in \autoref{fig:scalar}.

\begin{figure}[t]
	\includegraphics[width=\linewidth]{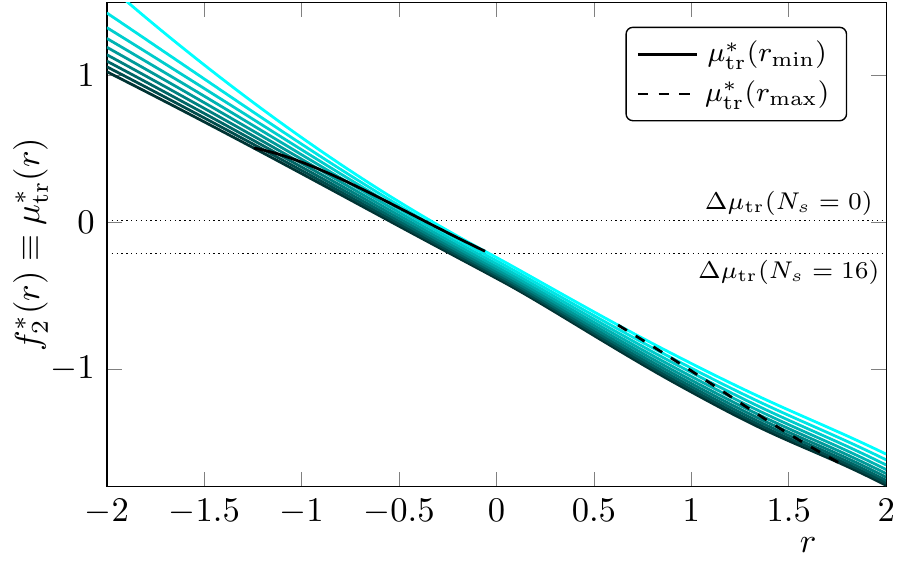}
	\caption{Stability of the solutions to the EoM. We included a shift in the $\mu_\text{tr}$ mode according to \eqref{eq:shift-mutr}.}
	\label{fig:stab-eom}
\end{figure}

The stability of a solution to the quantum EoM is described by the second fluctuation field derivative, in this case with respect to the trace mode.
The function $f_2$, which is defined in straight analogy to \eqref{eq:Gamma-f} and \eqref{eq:Gamma-f1},
is precisely the graviton mass parameter of the trace mode $\mu_\text{tr}(r)$. 
In our approximation, we have set $\mu_\text{tr} (r) = \mu_\text{tt}(r)$.
In \autoref{fig:stab-eom}, we display $\mu_\text{tr} (r)$.
We have a negative $\mu_\text{tr}$ for the solution at positive curvature, which is thus a maximum.
The solution at negative curvature starts negative but then becomes positive in this naive approximation.
This is an unphysical feature of our approximation and we expect this to be cured once the fully coupled system with $\mu_\text{tr}(r)$ is computed.
We test this by evaluating $\mu_\text{tr}(r=0)$ on the solution of the given fixed-point functions.
This allows us to improve our approximation and we can evaluate the trace mode by
\begin{align}
 \mu_\text{tr}(r) \approx \mu_\text{tr}(0) + \left[ \mu_\text{tt}(r) - \mu_\text{tt}(0) \right] \,.
 \label{eq:shift-mutr}
\end{align}
For $N_s=0$, we find that $\mu_\text{tr} (r) \approx \mu_\text{tt}(r)$,
while for $N_s=16$ they are shifted by $\mu_\text{tr} (r) \approx \mu_\text{tt}(r)-0.2$.
We included this shift in \autoref{fig:stab-eom} where now the solution at negative curvature is a minimum for all $N_s$.
It is actually remarkable that the two graviton mass parameters, $\mu_\text{tt}$ and $\mu_\text{tr}$, agree so well, in particular for small $N_s$, despite being related by non-trivial modified Slavnov-Taylor identities.
This is yet another sign of effective universality as introduced and discussed in \cite{Eichhorn:2018akn,Eichhorn:2018ydy}.

\subsection{Asymptotic behaviour and stability of scalar-gravity systems}
The functions $f$ and $f_1$ are not independent but are related by 
modified Slavnov-Taylor and Nielsen (or split Ward) identities
\cite{Litim:2002ce, Litim:2002hj,Pawlowski:2005xe, Folkerts:2011jz, Donkin:2012ud,
Bridle:2013sra, Dietz:2015owa, Safari:2015dva, Labus:2016lkh,Eichhorn:2018akn}. 
These identities carry $R_k$-dependent terms that reflect the additional $\bar g$-terms in $f(r)$ originating in the 
background dependence of the regulator. For $R_k\neq 0$ these identities are non-trivial. 
However, the right-hand sides of the flow equations for $f(r)$ and $f_1(r)$, \eqref{eq:Flowf} and \eqref{eq:Flowf1},
vanish for $| r| \to \infty$, i.e., for curvatures far bigger than the square of the cutoff scale, $k^2$, with the exception of possible zero modes. 
In this limit, the regulator is vanishing, $\lim_{r\to \infty} R_k(\bar g) \to 0$, and we are in the unregularised regime without cutoff effects.
Within the present approximation the effective action is diffeomorphism invariant for large curvatures, $\lim_{r\to \infty}\Gamma^*[\bar g, h]-S_\textrm{gf}= \Gamma^*[\bar g+h,0]-S_\textrm{gf}$, with vanishing ghosts.

The right-hand sides of the flow equations for $f(r)$ and $f_1(r)$ are vanishing for large curvatures 
and consequently, both, the background and the quantum EoM, asymptotically approach a solution.
Moreover, these solutions to the EoMs have to agree, $\bar{g}_{\overline{\text{EoM}}}=\bar{g}^{\ }_{{\text{EoM}}}$, 
due to the diffeomorphism invariance of the effective action for large curvatures.
This also entails that for $|r|\to\infty $ the fixed-point solutions $f_1^*$ and $f^*$ are related by a metric derivative, which leads to
\begin{align}
\label{eq:f-f1}
f^*_1(r) = r {f^*}'(r) -  2 f^* (r) \quad \text{for} \quad |r|\to \infty  \,. 
\end{align}
Now we use that the fixed-point solution of $f(r)$ is asymptotically either vanishing, $f(|r|\to\infty )= 0$, or $f(|r|\to\infty)\propto r^2$, see \eqref{eq:Flowf}.
In both cases is follows that $f_1(|r|\to \infty)=0$.
The differential equation \eqref{eq:Flowf1} also allows for the solution $f_1(|r|\to \infty)\propto |r|^{3/2}$, which is not compatible with the Nielsen identities.
Whether the fixed-point functions in \autoref{fig:qEoM} have the correct $f_1(|r|\to \infty)=0$ behaviour depends on $\mu_\text{eff}(r)$.
In turn, $\mu_\text{eff}(r)=f_2(r)$ is also constrained by a Nielsen identity, leading to $f^*_2(r) = r {f^*_1}'(r) -  2 f_1^* (r)$ for $|r|\to\infty$.
The differential equation for $\mu_\text{eff}$ allows again for two asymptotic solutions, $\mu_\text{eff}(|r|\to\infty )= 0$ or $\mu_\text{eff}(|r|\to\infty)\propto |r|$.
Only $\mu_\text{eff}(|r|\to\infty )= 0$ is compatible with the Nielsen identity. In conclusion, we can determine the asymptotic behaviour of the full tower of differential equations with the use of Nielsen identities.

With the latter properties, the fixed-point fluctuation Newton coupling $g^*(r)$ equals the background Newton coupling for $r\to\infty$.
They are positive and decay proportional to $1/r$ for positive curvature, see \autoref{fig:scalar}.
This entails a positive $r^2$ contribution in the background potential for large positive background curvature, $r/g^*(r) \propto  r^2$.
We determined the large curvature behaviour to be
\begin{align}
\label{eq:glarger}
\lim_{r\to \infty}g^*(r) &= \frac{c_+}{r}  \,,
\end{align}
with 
\begin{align}
c_+& = 0.92 - 8.6\cdot 10^{-3} N_s - 3.4 \cdot 10^{-4} N_s^2\,.
\end{align}
This behaviour was determined from the available fixed-point solutions with $N_s=0,\ldots,16$.
For $N_s>41$ this coefficient turns negative indicating the breakdown of the extrapolation.
The coefficient can never become negative since the flow equation for the Newton coupling has a Gau\ss ian fixed point.

In summary, this analysis leads us to the following global picture. 
For $|r|\leq 2$ we find two solutions of the quantum EoM, a minimum at negative curvature and a maximum at positive curvature.
For $N_s\approx17$ these solutions merge and disappear, see \autoref{fig:qEoM}.
For reasons of stability and due to the above considerations of the asymptotic behaviour,
we argue that at least one further solution to the EoM (a minimum) has to be present for $r> 2$.
While it might not be the absolute minimum in pure quantum gravity and small $N_s$, it is for large enough $N_s$ as  the other minimum disappears.
Importantly, for large positive curvature, the fixed point has a nice stable behaviour and the Newton coupling remains small for large $N_s$, see \autoref{fig:scalar}.
This leads to the important and exciting conclusion that the dynamics of gravity stabilises the fixed point and effectively dominates over the matter fluctuations.
This has first been seen and discussed in \cite{Meibohm:2015twa} and extended in \cite{Christiansen:2017cxa}: \textit{one force to rule them all}.
In \cite{Meibohm:2015twa} the respective behaviour for fermions with a decreasing Newton coupling was already seen at $r=0$.
In turn, for scalars, this behaviour was not seen at $r=0$ and the approximation breaks down for large $N_s$ due to the increasing fixed-point value of the Newton coupling.
The present scenario hints at a \textit{geometrical} first-order phase transition in the curvature with the number of scalars.
This scenario will be investigated in more details elsewhere.  

\section{Summary and Outlook}
\label{sec:summary}
In the present work, we have discussed fixed-point solutions of quantum gravity including an $f(r)$-term for constant background curvatures $r\in\mathbb{R}$ with minimally coupled scalar fields.
This extends the work initiated in \cite{Christiansen:2017bsy} to negative curvatures, $r< 0$.
In \cite{Christiansen:2017bsy}, a novel expansion scheme was put forward that allows for the computation of (background) curvature-dependent correlation functions of the fluctuation field. 
This gives access to the diffeomorphism invariant background effective action and, in particular, the $f(r)$ potential without resorting to the background-field approximation. 

We found fixed-point functions for the couplings of the graviton two- and three-point functions, $g(r)$, $\mu(r)$ and $\lambda_3(r)$, up to $N_s=16$.
For larger $N_s$, there is a divergence in the Newton coupling at negative background curvature.
This divergence is not a physical feature but signals the breakdown of the truncation.
Indeed, we found that the system is remarkably stable for positive background curvature.
We emphasise that this can be seen as an indication for the persistence of the asymptotically safe UV fixed point for $N_s\to\infty$.

We have discussed the solution to the background and quantum EoM for curvatures $|r| \lesssim 2$ as well as their asymptotic behaviour.
The background EoM has no solution in this curvature regime for any $N_s$.
The quantum EoM has two solutions for $N_s\leq 16$ in this regime, a minimum at negative background curvature and a maximum at positive background curvature. 
We have argued that the global structure of the fixed-point solution should admit a second minimum at large positive curvatures also for large $N_s$.
Since the fixed-point solutions are well behaved in this curvature regime and the Newton coupling remains small,
this stabilises asymptotically safe gravity for a large number of scalars in line with the mechanism introduced in \cite{Meibohm:2015twa, Christiansen:2017cxa}: \textit{one force to rule them all}.
The disappearance of the minimum at negative curvature can be interpreted as a hint for a geometrical first-order phase transition with the number of scalars.

In summary, together with earlier results on matter-gravity systems with only gravitationally interacting matter, also including fermions and gauge bosons, 
we conclude that there are indications for the global stability of gravity-matter systems.
This is an important result as, for the first time, we can pinpoint why previous works \cite{Dona:2013qba,Meibohm:2015twa,Dona:2015tnf,Eichhorn:2018akn,Alkofer:2018fxj} have observed a seeming divergence in the Newton coupling with increasing $N_s$.
With an expansion about the minimum at large positive curvature these systems should show the stability properties that have been derived from the path-integral representation in \cite{Christiansen:2017cxa}. 
This now allows for a reliable discussion of the stability of gravity-matter systems and phenomenological high energy and phenomenological applications such as \cite{Eichhorn:2017ylw,Eichhorn:2018whv,Reichert:2019car}.

\vspace{.3cm}
\noindent {\bf Acknowledgements} 
This work is supported by the Danish National Research Foundation under grant DNRF:90. It is part of and supported by the DFG
Collaborative Research Centre SFB 1225 (ISOQUANT) as well as by the
DFG under Germany's Excellence Strategy EXC - 2181/1 - 390900948 (the
Heidelberg Excellence Cluster STRUCTURES). 


\bibliography{flatgravity}

\begin{thebibliography}{78}%
\makeatletter
\providecommand \@ifxundefined [1]{%
 \@ifx{#1\undefined}
}%
\providecommand \@ifnum [1]{%
 \ifnum #1\expandafter \@firstoftwo
 \else \expandafter \@secondoftwo
 \fi
}%
\providecommand \@ifx [1]{%
 \ifx #1\expandafter \@firstoftwo
 \else \expandafter \@secondoftwo
 \fi
}%
\providecommand \natexlab [1]{#1}%
\providecommand \enquote  [1]{``#1''}%
\providecommand \bibnamefont  [1]{#1}%
\providecommand \bibfnamefont [1]{#1}%
\providecommand \citenamefont [1]{#1}%
\providecommand \href@noop [0]{\@secondoftwo}%
\providecommand \href [0]{\begingroup \@sanitize@url \@href}%
\providecommand \@href[1]{\@@startlink{#1}\@@href}%
\providecommand \@@href[1]{\endgroup#1\@@endlink}%
\providecommand \@sanitize@url [0]{\catcode `\\12\catcode `\$12\catcode
  `\&12\catcode `\#12\catcode `\^12\catcode `\_12\catcode `\%12\relax}%
\providecommand \@@startlink[1]{}%
\providecommand \@@endlink[0]{}%
\providecommand \url  [0]{\begingroup\@sanitize@url \@url }%
\providecommand \@url [1]{\endgroup\@href {#1}{\urlprefix }}%
\providecommand \urlprefix  [0]{URL }%
\providecommand \Eprint [0]{\href }%
\providecommand \doibase [0]{http://dx.doi.org/}%
\providecommand \selectlanguage [0]{\@gobble}%
\providecommand \bibinfo  [0]{\@secondoftwo}%
\providecommand \bibfield  [0]{\@secondoftwo}%
\providecommand \translation [1]{[#1]}%
\providecommand \BibitemOpen [0]{}%
\providecommand \bibitemStop [0]{}%
\providecommand \bibitemNoStop [0]{.\EOS\space}%
\providecommand \EOS [0]{\spacefactor3000\relax}%
\providecommand \BibitemShut  [1]{\csname bibitem#1\endcsname}%
\let\auto@bib@innerbib\@empty
\bibitem [{\citenamefont {Christiansen}\ \emph
  {et~al.}(2018{\natexlab{a}})\citenamefont {Christiansen}, \citenamefont
  {Falls}, \citenamefont {Pawlowski},\ and\ \citenamefont
  {Reichert}}]{Christiansen:2017bsy}%
  \BibitemOpen
  \bibfield  {author} {\bibinfo {author} {\bibfnamefont {N.}~\bibnamefont
  {Christiansen}}, \bibinfo {author} {\bibfnamefont {K.}~\bibnamefont {Falls}},
  \bibinfo {author} {\bibfnamefont {J.~M.}\ \bibnamefont {Pawlowski}}, \ and\
  \bibinfo {author} {\bibfnamefont {M.}~\bibnamefont {Reichert}},\ }\href
  {\doibase 10.1103/PhysRevD.97.046007} {\bibfield  {journal} {\bibinfo
  {journal} {Phys. Rev.}\ }\textbf {\bibinfo {volume} {D97}},\ \bibinfo {pages}
  {046007} (\bibinfo {year} {2018}{\natexlab{a}})},\ \Eprint
  {http://arxiv.org/abs/1711.09259} {arXiv:1711.09259 [hep-th]} \BibitemShut
  {NoStop}%
\bibitem [{\citenamefont {Weinberg}(1979)}]{Weinberg:1980gg}%
  \BibitemOpen
  \bibfield  {author} {\bibinfo {author} {\bibfnamefont {S.}~\bibnamefont
  {Weinberg}},\ }\href@noop {} {\bibfield  {journal} {\bibinfo  {journal}
  {General Relativity: An Einstein centenary survey, Eds. Hawking, S.W.,
  Israel, W; Cambridge University Press}\ ,\ \bibinfo {pages} {790}} (\bibinfo
  {year} {1979})}\BibitemShut {NoStop}%
\bibitem [{\citenamefont {Reuter}(1998)}]{Reuter:1996cp}%
  \BibitemOpen
  \bibfield  {author} {\bibinfo {author} {\bibfnamefont {M.}~\bibnamefont
  {Reuter}},\ }\href {\doibase 10.1103/PhysRevD.57.971} {\bibfield  {journal}
  {\bibinfo  {journal} {Phys. Rev.}\ }\textbf {\bibinfo {volume} {D57}},\
  \bibinfo {pages} {971} (\bibinfo {year} {1998})},\ \Eprint
  {http://arxiv.org/abs/hep-th/9605030} {arXiv:hep-th/9605030} \BibitemShut
  {NoStop}%
\bibitem [{\citenamefont {Niedermaier}\ and\ \citenamefont
  {Reuter}(2006)}]{Niedermaier:2006wt}%
  \BibitemOpen
  \bibfield  {author} {\bibinfo {author} {\bibfnamefont {M.}~\bibnamefont
  {Niedermaier}}\ and\ \bibinfo {author} {\bibfnamefont {M.}~\bibnamefont
  {Reuter}},\ }\href@noop {} {\bibfield  {journal} {\bibinfo  {journal} {Living
  Rev.Rel.}\ }\textbf {\bibinfo {volume} {9}},\ \bibinfo {pages} {5} (\bibinfo
  {year} {2006})}\BibitemShut {NoStop}%
\bibitem [{\citenamefont {Litim}(2011)}]{Litim:2011cp}%
  \BibitemOpen
  \bibfield  {author} {\bibinfo {author} {\bibfnamefont {D.~F.}\ \bibnamefont
  {Litim}},\ }\href@noop {} {\bibfield  {journal} {\bibinfo  {journal}
  {Phil.Trans.Roy.Soc.Lond.}\ }\textbf {\bibinfo {volume} {A369}},\ \bibinfo
  {pages} {2759} (\bibinfo {year} {2011})},\ \Eprint
  {http://arxiv.org/abs/1102.4624} {arXiv:1102.4624 [hep-th]} \BibitemShut
  {NoStop}%
\bibitem [{\citenamefont {Reuter}\ and\ \citenamefont
  {Saueressig}(2012)}]{Reuter:2012id}%
  \BibitemOpen
  \bibfield  {author} {\bibinfo {author} {\bibfnamefont {M.}~\bibnamefont
  {Reuter}}\ and\ \bibinfo {author} {\bibfnamefont {F.}~\bibnamefont
  {Saueressig}},\ }\href {\doibase 10.1088/1367-2630/14/5/055022} {\bibfield
  {journal} {\bibinfo  {journal} {New J. Phys.}\ }\textbf {\bibinfo {volume}
  {14}},\ \bibinfo {pages} {055022} (\bibinfo {year} {2012})},\ \Eprint
  {http://arxiv.org/abs/1202.2274} {arXiv:1202.2274 [hep-th]} \BibitemShut
  {NoStop}%
\bibitem [{\citenamefont {Bonanno}\ and\ \citenamefont
  {Saueressig}(2017)}]{Bonanno:2017pkg}%
  \BibitemOpen
  \bibfield  {author} {\bibinfo {author} {\bibfnamefont {A.}~\bibnamefont
  {Bonanno}}\ and\ \bibinfo {author} {\bibfnamefont {F.}~\bibnamefont
  {Saueressig}},\ }\href {\doibase 10.1016/j.crhy.2017.02.002} {\bibfield
  {journal} {\bibinfo  {journal} {Comptes Rendus Physique}\ }\textbf {\bibinfo
  {volume} {18}},\ \bibinfo {pages} {254} (\bibinfo {year} {2017})},\ \Eprint
  {http://arxiv.org/abs/1702.04137} {arXiv:1702.04137 [hep-th]} \BibitemShut
  {NoStop}%
\bibitem [{\citenamefont {Eichhorn}(2017)}]{Eichhorn:2017egq}%
  \BibitemOpen
  \bibfield  {author} {\bibinfo {author} {\bibfnamefont {A.}~\bibnamefont
  {Eichhorn}},\ }in\ \href@noop {} {\emph {\bibinfo {booktitle} {{Black Holes,
  Gravitational Waves and Spacetime Singularities Rome, Italy, May 9-12,
  2017}}}}\ (\bibinfo {year} {2017})\ \Eprint {http://arxiv.org/abs/1709.03696}
  {arXiv:1709.03696 [gr-qc]} \BibitemShut {NoStop}%
\bibitem [{\citenamefont {Percacci}(2017)}]{Percacci:2017fkn}%
  \BibitemOpen
  \bibfield  {author} {\bibinfo {author} {\bibfnamefont {R.}~\bibnamefont
  {Percacci}},\ }\href {\doibase 10.1142/10369} {\emph {\bibinfo {title} {{An
  Introduction to Covariant Quantum Gravity and Asymptotic Safety}}}},\
  \bibinfo {series} {{100 Years of General Relativity}}, Vol.~\bibinfo {volume}
  {3}\ (\bibinfo  {publisher} {World Scientific},\ \bibinfo {year}
  {2017})\BibitemShut {NoStop}%
\bibitem [{\citenamefont {Wetterich}(2019{\natexlab{a}})}]{Wetterich:2019qzx}%
  \BibitemOpen
  \bibfield  {author} {\bibinfo {author} {\bibfnamefont {C.}~\bibnamefont
  {Wetterich}},\ }\href@noop {} {\  (\bibinfo {year} {2019}{\natexlab{a}})},\
  \Eprint {http://arxiv.org/abs/1901.04741} {arXiv:1901.04741 [hep-th]}
  \BibitemShut {NoStop}%
\bibitem [{\citenamefont {Wetterich}(1993)}]{Wetterich:1992yh}%
  \BibitemOpen
  \bibfield  {author} {\bibinfo {author} {\bibfnamefont {C.}~\bibnamefont
  {Wetterich}},\ }\href {\doibase 10.1016/0370-2693(93)90726-X} {\bibfield
  {journal} {\bibinfo  {journal} {Phys. Lett.}\ }\textbf {\bibinfo {volume}
  {B301}},\ \bibinfo {pages} {90} (\bibinfo {year} {1993})},\ \Eprint
  {http://arxiv.org/abs/1710.05815} {arXiv:1710.05815 [hep-th]} \BibitemShut
  {NoStop}%
\bibitem [{\citenamefont {Christiansen}\ \emph {et~al.}(2014)\citenamefont
  {Christiansen}, \citenamefont {Litim}, \citenamefont {Pawlowski},\ and\
  \citenamefont {Rodigast}}]{Christiansen:2012rx}%
  \BibitemOpen
  \bibfield  {author} {\bibinfo {author} {\bibfnamefont {N.}~\bibnamefont
  {Christiansen}}, \bibinfo {author} {\bibfnamefont {D.~F.}\ \bibnamefont
  {Litim}}, \bibinfo {author} {\bibfnamefont {J.~M.}\ \bibnamefont
  {Pawlowski}}, \ and\ \bibinfo {author} {\bibfnamefont {A.}~\bibnamefont
  {Rodigast}},\ }\href {\doibase 10.1016/j.physletb.2013.11.025} {\bibfield
  {journal} {\bibinfo  {journal} {Phys.Lett.}\ }\textbf {\bibinfo {volume}
  {B728}},\ \bibinfo {pages} {114} (\bibinfo {year} {2014})},\ \Eprint
  {http://arxiv.org/abs/1209.4038} {arXiv:1209.4038 [hep-th]} \BibitemShut
  {NoStop}%
\bibitem [{\citenamefont {Christiansen}\ \emph {et~al.}(2016)\citenamefont
  {Christiansen}, \citenamefont {Knorr}, \citenamefont {Pawlowski},\ and\
  \citenamefont {Rodigast}}]{Christiansen:2014raa}%
  \BibitemOpen
  \bibfield  {author} {\bibinfo {author} {\bibfnamefont {N.}~\bibnamefont
  {Christiansen}}, \bibinfo {author} {\bibfnamefont {B.}~\bibnamefont {Knorr}},
  \bibinfo {author} {\bibfnamefont {J.~M.}\ \bibnamefont {Pawlowski}}, \ and\
  \bibinfo {author} {\bibfnamefont {A.}~\bibnamefont {Rodigast}},\ }\href
  {\doibase 10.1103/PhysRevD.93.044036} {\bibfield  {journal} {\bibinfo
  {journal} {Phys. Rev.}\ }\textbf {\bibinfo {volume} {D93}},\ \bibinfo {pages}
  {044036} (\bibinfo {year} {2016})},\ \Eprint {http://arxiv.org/abs/1403.1232}
  {arXiv:1403.1232 [hep-th]} \BibitemShut {NoStop}%
\bibitem [{\citenamefont {Christiansen}\ \emph {et~al.}(2015)\citenamefont
  {Christiansen}, \citenamefont {Knorr}, \citenamefont {Meibohm}, \citenamefont
  {Pawlowski},\ and\ \citenamefont {Reichert}}]{Christiansen:2015rva}%
  \BibitemOpen
  \bibfield  {author} {\bibinfo {author} {\bibfnamefont {N.}~\bibnamefont
  {Christiansen}}, \bibinfo {author} {\bibfnamefont {B.}~\bibnamefont {Knorr}},
  \bibinfo {author} {\bibfnamefont {J.}~\bibnamefont {Meibohm}}, \bibinfo
  {author} {\bibfnamefont {J.~M.}\ \bibnamefont {Pawlowski}}, \ and\ \bibinfo
  {author} {\bibfnamefont {M.}~\bibnamefont {Reichert}},\ }\href {\doibase
  10.1103/PhysRevD.92.121501} {\bibfield  {journal} {\bibinfo  {journal} {Phys.
  Rev.}\ }\textbf {\bibinfo {volume} {D92}},\ \bibinfo {pages} {121501}
  (\bibinfo {year} {2015})},\ \Eprint {http://arxiv.org/abs/1506.07016}
  {arXiv:1506.07016 [hep-th]} \BibitemShut {NoStop}%
\bibitem [{\citenamefont {Christiansen}(2016)}]{Christiansen:2016sjn}%
  \BibitemOpen
  \bibfield  {author} {\bibinfo {author} {\bibfnamefont {N.}~\bibnamefont
  {Christiansen}},\ }\href@noop {} {\  (\bibinfo {year} {2016})},\ \Eprint
  {http://arxiv.org/abs/1612.06223} {arXiv:1612.06223 [hep-th]} \BibitemShut
  {NoStop}%
\bibitem [{\citenamefont {Denz}\ \emph {et~al.}(2018)\citenamefont {Denz},
  \citenamefont {Pawlowski},\ and\ \citenamefont {Reichert}}]{Denz:2016qks}%
  \BibitemOpen
  \bibfield  {author} {\bibinfo {author} {\bibfnamefont {T.}~\bibnamefont
  {Denz}}, \bibinfo {author} {\bibfnamefont {J.~M.}\ \bibnamefont {Pawlowski}},
  \ and\ \bibinfo {author} {\bibfnamefont {M.}~\bibnamefont {Reichert}},\
  }\href {\doibase 10.1140/epjc/s10052-018-5806-0} {\bibfield  {journal}
  {\bibinfo  {journal} {Eur. Phys. J.}\ }\textbf {\bibinfo {volume} {C78}},\
  \bibinfo {pages} {336} (\bibinfo {year} {2018})},\ \Eprint
  {http://arxiv.org/abs/1612.07315} {arXiv:1612.07315 [hep-th]} \BibitemShut
  {NoStop}%
\bibitem [{\citenamefont {Eichhorn}\ \emph
  {et~al.}(2019{\natexlab{a}})\citenamefont {Eichhorn}, \citenamefont
  {Lippoldt}, \citenamefont {Pawlowski}, \citenamefont {Reichert},\ and\
  \citenamefont {Schiffer}}]{Eichhorn:2018ydy}%
  \BibitemOpen
  \bibfield  {author} {\bibinfo {author} {\bibfnamefont {A.}~\bibnamefont
  {Eichhorn}}, \bibinfo {author} {\bibfnamefont {S.}~\bibnamefont {Lippoldt}},
  \bibinfo {author} {\bibfnamefont {J.~M.}\ \bibnamefont {Pawlowski}}, \bibinfo
  {author} {\bibfnamefont {M.}~\bibnamefont {Reichert}}, \ and\ \bibinfo
  {author} {\bibfnamefont {M.}~\bibnamefont {Schiffer}},\ }\href {\doibase
  10.1016/j.physletb.2019.01.071} {\bibfield  {journal} {\bibinfo  {journal}
  {Phys. Lett.}\ }\textbf {\bibinfo {volume} {B792}},\ \bibinfo {pages} {310}
  (\bibinfo {year} {2019}{\natexlab{a}})},\ \Eprint
  {http://arxiv.org/abs/1810.02828} {arXiv:1810.02828 [hep-th]} \BibitemShut
  {NoStop}%
\bibitem [{\citenamefont {Eichhorn}\ \emph
  {et~al.}(2019{\natexlab{b}})\citenamefont {Eichhorn}, \citenamefont
  {Lippoldt},\ and\ \citenamefont {Schiffer}}]{Eichhorn:2018nda}%
  \BibitemOpen
  \bibfield  {author} {\bibinfo {author} {\bibfnamefont {A.}~\bibnamefont
  {Eichhorn}}, \bibinfo {author} {\bibfnamefont {S.}~\bibnamefont {Lippoldt}},
  \ and\ \bibinfo {author} {\bibfnamefont {M.}~\bibnamefont {Schiffer}},\
  }\href {\doibase 10.1103/PhysRevD.99.086002} {\bibfield  {journal} {\bibinfo
  {journal} {Phys. Rev.}\ }\textbf {\bibinfo {volume} {D99}},\ \bibinfo {pages}
  {086002} (\bibinfo {year} {2019}{\natexlab{b}})},\ \Eprint
  {http://arxiv.org/abs/1812.08782} {arXiv:1812.08782 [hep-th]} \BibitemShut
  {NoStop}%
\bibitem [{\citenamefont {Knorr}\ and\ \citenamefont
  {Lippoldt}(2017)}]{Knorr:2017fus}%
  \BibitemOpen
  \bibfield  {author} {\bibinfo {author} {\bibfnamefont {B.}~\bibnamefont
  {Knorr}}\ and\ \bibinfo {author} {\bibfnamefont {S.}~\bibnamefont
  {Lippoldt}},\ }\href {\doibase 10.1103/PhysRevD.96.065020} {\bibfield
  {journal} {\bibinfo  {journal} {Phys. Rev.}\ }\textbf {\bibinfo {volume}
  {D96}},\ \bibinfo {pages} {065020} (\bibinfo {year} {2017})},\ \Eprint
  {http://arxiv.org/abs/1707.01397} {arXiv:1707.01397 [hep-th]} \BibitemShut
  {NoStop}%
\bibitem [{\citenamefont {Knorr}(2018)}]{Knorr:2017mhu}%
  \BibitemOpen
  \bibfield  {author} {\bibinfo {author} {\bibfnamefont {B.}~\bibnamefont
  {Knorr}},\ }\href {\doibase 10.1088/1361-6382/aabaa0} {\bibfield  {journal}
  {\bibinfo  {journal} {Class. Quant. Grav.}\ }\textbf {\bibinfo {volume}
  {35}},\ \bibinfo {pages} {115005} (\bibinfo {year} {2018})},\ \Eprint
  {http://arxiv.org/abs/1710.07055} {arXiv:1710.07055 [hep-th]} \BibitemShut
  {NoStop}%
\bibitem [{\citenamefont {Codello}\ \emph {et~al.}(2008)\citenamefont
  {Codello}, \citenamefont {Percacci},\ and\ \citenamefont
  {Rahmede}}]{Codello:2007bd}%
  \BibitemOpen
  \bibfield  {author} {\bibinfo {author} {\bibfnamefont {A.}~\bibnamefont
  {Codello}}, \bibinfo {author} {\bibfnamefont {R.}~\bibnamefont {Percacci}}, \
  and\ \bibinfo {author} {\bibfnamefont {C.}~\bibnamefont {Rahmede}},\ }\href
  {\doibase 10.1142/S0217751X08038135} {\bibfield  {journal} {\bibinfo
  {journal} {Int. J. Mod. Phys.}\ }\textbf {\bibinfo {volume} {A23}},\ \bibinfo
  {pages} {143} (\bibinfo {year} {2008})},\ \Eprint
  {http://arxiv.org/abs/0705.1769} {arXiv:0705.1769 [hep-th]} \BibitemShut
  {NoStop}%
\bibitem [{\citenamefont {Machado}\ and\ \citenamefont
  {Saueressig}(2008)}]{Machado:2007ea}%
  \BibitemOpen
  \bibfield  {author} {\bibinfo {author} {\bibfnamefont {P.~F.}\ \bibnamefont
  {Machado}}\ and\ \bibinfo {author} {\bibfnamefont {F.}~\bibnamefont
  {Saueressig}},\ }\href {\doibase 10.1103/PhysRevD.77.124045} {\bibfield
  {journal} {\bibinfo  {journal} {Phys. Rev.}\ }\textbf {\bibinfo {volume}
  {D77}},\ \bibinfo {pages} {124045} (\bibinfo {year} {2008})},\ \Eprint
  {http://arxiv.org/abs/0712.0445} {arXiv:0712.0445 [hep-th]} \BibitemShut
  {NoStop}%
\bibitem [{\citenamefont {Codello}\ \emph {et~al.}(2009)\citenamefont
  {Codello}, \citenamefont {Percacci},\ and\ \citenamefont
  {Rahmede}}]{Codello:2008vh}%
  \BibitemOpen
  \bibfield  {author} {\bibinfo {author} {\bibfnamefont {A.}~\bibnamefont
  {Codello}}, \bibinfo {author} {\bibfnamefont {R.}~\bibnamefont {Percacci}}, \
  and\ \bibinfo {author} {\bibfnamefont {C.}~\bibnamefont {Rahmede}},\ }\href
  {\doibase 10.1016/j.aop.2008.08.008} {\bibfield  {journal} {\bibinfo
  {journal} {Annals Phys.}\ }\textbf {\bibinfo {volume} {324}},\ \bibinfo
  {pages} {414} (\bibinfo {year} {2009})},\ \Eprint
  {http://arxiv.org/abs/0805.2909} {arXiv:0805.2909 [hep-th]} \BibitemShut
  {NoStop}%
\bibitem [{\citenamefont {Benedetti}\ and\ \citenamefont
  {Caravelli}(2012)}]{Benedetti:2012dx}%
  \BibitemOpen
  \bibfield  {author} {\bibinfo {author} {\bibfnamefont {D.}~\bibnamefont
  {Benedetti}}\ and\ \bibinfo {author} {\bibfnamefont {F.}~\bibnamefont
  {Caravelli}},\ }\href {\doibase
  10.1007/JHEP06(2012)017;;;;;;;;;;;;;;;;;;;;;;;;;;;;;;;;;;
  10.1007/JHEP10(2012)157} {\bibfield  {journal} {\bibinfo  {journal} {JHEP}\
  }\textbf {\bibinfo {volume} {06}},\ \bibinfo {pages} {017} (\bibinfo {year}
  {2012})},\ \bibinfo {note} {[Erratum: JHEP10,157(2012)]},\ \Eprint
  {http://arxiv.org/abs/1204.3541} {arXiv:1204.3541 [hep-th]} \BibitemShut
  {NoStop}%
\bibitem [{\citenamefont {Dietz}\ and\ \citenamefont
  {Morris}(2013{\natexlab{a}})}]{Dietz:2012ic}%
  \BibitemOpen
  \bibfield  {author} {\bibinfo {author} {\bibfnamefont {J.~A.}\ \bibnamefont
  {Dietz}}\ and\ \bibinfo {author} {\bibfnamefont {T.~R.}\ \bibnamefont
  {Morris}},\ }\href {\doibase 10.1007/JHEP01(2013)108} {\bibfield  {journal}
  {\bibinfo  {journal} {JHEP}\ }\textbf {\bibinfo {volume} {01}},\ \bibinfo
  {pages} {108} (\bibinfo {year} {2013}{\natexlab{a}})},\ \Eprint
  {http://arxiv.org/abs/1211.0955} {arXiv:1211.0955 [hep-th]} \BibitemShut
  {NoStop}%
\bibitem [{\citenamefont {Falls}\ \emph {et~al.}(2013)\citenamefont {Falls},
  \citenamefont {Litim}, \citenamefont {Nikolakopoulos},\ and\ \citenamefont
  {Rahmede}}]{Falls:2013bv}%
  \BibitemOpen
  \bibfield  {author} {\bibinfo {author} {\bibfnamefont {K.}~\bibnamefont
  {Falls}}, \bibinfo {author} {\bibfnamefont {D.}~\bibnamefont {Litim}},
  \bibinfo {author} {\bibfnamefont {K.}~\bibnamefont {Nikolakopoulos}}, \ and\
  \bibinfo {author} {\bibfnamefont {C.}~\bibnamefont {Rahmede}},\ }\href@noop
  {} {\  (\bibinfo {year} {2013})},\ \Eprint {http://arxiv.org/abs/1301.4191}
  {arXiv:1301.4191 [hep-th]} \BibitemShut {NoStop}%
\bibitem [{\citenamefont {Benedetti}(2013)}]{Benedetti:2013jk}%
  \BibitemOpen
  \bibfield  {author} {\bibinfo {author} {\bibfnamefont {D.}~\bibnamefont
  {Benedetti}},\ }\href {\doibase 10.1209/0295-5075/102/20007} {\bibfield
  {journal} {\bibinfo  {journal} {Europhys. Lett.}\ }\textbf {\bibinfo {volume}
  {102}},\ \bibinfo {pages} {20007} (\bibinfo {year} {2013})},\ \Eprint
  {http://arxiv.org/abs/1301.4422} {arXiv:1301.4422 [hep-th]} \BibitemShut
  {NoStop}%
\bibitem [{\citenamefont {Dietz}\ and\ \citenamefont
  {Morris}(2013{\natexlab{b}})}]{Dietz:2013sba}%
  \BibitemOpen
  \bibfield  {author} {\bibinfo {author} {\bibfnamefont {J.~A.}\ \bibnamefont
  {Dietz}}\ and\ \bibinfo {author} {\bibfnamefont {T.~R.}\ \bibnamefont
  {Morris}},\ }\href {\doibase 10.1007/JHEP07(2013)064} {\bibfield  {journal}
  {\bibinfo  {journal} {JHEP}\ }\textbf {\bibinfo {volume} {07}},\ \bibinfo
  {pages} {064} (\bibinfo {year} {2013}{\natexlab{b}})},\ \Eprint
  {http://arxiv.org/abs/1306.1223} {arXiv:1306.1223 [hep-th]} \BibitemShut
  {NoStop}%
\bibitem [{\citenamefont {Demmel}\ \emph {et~al.}(2014)\citenamefont {Demmel},
  \citenamefont {Saueressig},\ and\ \citenamefont {Zanusso}}]{Demmel:2014sga}%
  \BibitemOpen
  \bibfield  {author} {\bibinfo {author} {\bibfnamefont {M.}~\bibnamefont
  {Demmel}}, \bibinfo {author} {\bibfnamefont {F.}~\bibnamefont {Saueressig}},
  \ and\ \bibinfo {author} {\bibfnamefont {O.}~\bibnamefont {Zanusso}},\ }\href
  {\doibase 10.1007/JHEP06(2014)026} {\bibfield  {journal} {\bibinfo  {journal}
  {JHEP}\ }\textbf {\bibinfo {volume} {06}},\ \bibinfo {pages} {026} (\bibinfo
  {year} {2014})},\ \Eprint {http://arxiv.org/abs/1401.5495} {arXiv:1401.5495
  [hep-th]} \BibitemShut {NoStop}%
\bibitem [{\citenamefont {Falls}\ \emph {et~al.}(2016)\citenamefont {Falls},
  \citenamefont {Litim}, \citenamefont {Nikolakopoulos},\ and\ \citenamefont
  {Rahmede}}]{Falls:2014tra}%
  \BibitemOpen
  \bibfield  {author} {\bibinfo {author} {\bibfnamefont {K.}~\bibnamefont
  {Falls}}, \bibinfo {author} {\bibfnamefont {D.~F.}\ \bibnamefont {Litim}},
  \bibinfo {author} {\bibfnamefont {K.}~\bibnamefont {Nikolakopoulos}}, \ and\
  \bibinfo {author} {\bibfnamefont {C.}~\bibnamefont {Rahmede}},\ }\href
  {\doibase 10.1103/PhysRevD.93.104022} {\bibfield  {journal} {\bibinfo
  {journal} {Phys. Rev.}\ }\textbf {\bibinfo {volume} {D93}},\ \bibinfo {pages}
  {104022} (\bibinfo {year} {2016})},\ \Eprint {http://arxiv.org/abs/1410.4815}
  {arXiv:1410.4815 [hep-th]} \BibitemShut {NoStop}%
\bibitem [{\citenamefont {Demmel}\ \emph {et~al.}(2015)\citenamefont {Demmel},
  \citenamefont {Saueressig},\ and\ \citenamefont {Zanusso}}]{Demmel:2015oqa}%
  \BibitemOpen
  \bibfield  {author} {\bibinfo {author} {\bibfnamefont {M.}~\bibnamefont
  {Demmel}}, \bibinfo {author} {\bibfnamefont {F.}~\bibnamefont {Saueressig}},
  \ and\ \bibinfo {author} {\bibfnamefont {O.}~\bibnamefont {Zanusso}},\ }\href
  {\doibase 10.1007/JHEP08(2015)113} {\bibfield  {journal} {\bibinfo  {journal}
  {JHEP}\ }\textbf {\bibinfo {volume} {08}},\ \bibinfo {pages} {113} (\bibinfo
  {year} {2015})},\ \Eprint {http://arxiv.org/abs/1504.07656} {arXiv:1504.07656
  [hep-th]} \BibitemShut {NoStop}%
\bibitem [{\citenamefont {Ohta}\ \emph {et~al.}(2015)\citenamefont {Ohta},
  \citenamefont {Percacci},\ and\ \citenamefont {Vacca}}]{Ohta:2015efa}%
  \BibitemOpen
  \bibfield  {author} {\bibinfo {author} {\bibfnamefont {N.}~\bibnamefont
  {Ohta}}, \bibinfo {author} {\bibfnamefont {R.}~\bibnamefont {Percacci}}, \
  and\ \bibinfo {author} {\bibfnamefont {G.~P.}\ \bibnamefont {Vacca}},\ }\href
  {\doibase 10.1103/PhysRevD.92.061501} {\bibfield  {journal} {\bibinfo
  {journal} {Phys. Rev.}\ }\textbf {\bibinfo {volume} {D92}},\ \bibinfo {pages}
  {061501} (\bibinfo {year} {2015})},\ \Eprint
  {http://arxiv.org/abs/1507.00968} {arXiv:1507.00968 [hep-th]} \BibitemShut
  {NoStop}%
\bibitem [{\citenamefont {Ohta}\ \emph {et~al.}(2016)\citenamefont {Ohta},
  \citenamefont {Percacci},\ and\ \citenamefont {Vacca}}]{Ohta:2015fcu}%
  \BibitemOpen
  \bibfield  {author} {\bibinfo {author} {\bibfnamefont {N.}~\bibnamefont
  {Ohta}}, \bibinfo {author} {\bibfnamefont {R.}~\bibnamefont {Percacci}}, \
  and\ \bibinfo {author} {\bibfnamefont {G.~P.}\ \bibnamefont {Vacca}},\ }\href
  {\doibase 10.1140/epjc/s10052-016-3895-1} {\bibfield  {journal} {\bibinfo
  {journal} {Eur. Phys. J.}\ }\textbf {\bibinfo {volume} {C76}},\ \bibinfo
  {pages} {46} (\bibinfo {year} {2016})},\ \Eprint
  {http://arxiv.org/abs/1511.09393} {arXiv:1511.09393 [hep-th]} \BibitemShut
  {NoStop}%
\bibitem [{\citenamefont {Falls}\ \emph
  {et~al.}(2018{\natexlab{a}})\citenamefont {Falls}, \citenamefont {Litim},
  \citenamefont {Nikolakopoulos},\ and\ \citenamefont
  {Rahmede}}]{Falls:2016wsa}%
  \BibitemOpen
  \bibfield  {author} {\bibinfo {author} {\bibfnamefont {K.}~\bibnamefont
  {Falls}}, \bibinfo {author} {\bibfnamefont {D.~F.}\ \bibnamefont {Litim}},
  \bibinfo {author} {\bibfnamefont {K.}~\bibnamefont {Nikolakopoulos}}, \ and\
  \bibinfo {author} {\bibfnamefont {C.}~\bibnamefont {Rahmede}},\ }\href
  {\doibase 10.1088/1361-6382/aac440} {\bibfield  {journal} {\bibinfo
  {journal} {Class. Quant. Grav.}\ }\textbf {\bibinfo {volume} {35}},\ \bibinfo
  {pages} {135006} (\bibinfo {year} {2018}{\natexlab{a}})},\ \Eprint
  {http://arxiv.org/abs/1607.04962} {arXiv:1607.04962 [gr-qc]} \BibitemShut
  {NoStop}%
\bibitem [{\citenamefont {Falls}\ and\ \citenamefont
  {Ohta}(2016)}]{Falls:2016msz}%
  \BibitemOpen
  \bibfield  {author} {\bibinfo {author} {\bibfnamefont {K.}~\bibnamefont
  {Falls}}\ and\ \bibinfo {author} {\bibfnamefont {N.}~\bibnamefont {Ohta}},\
  }\href {\doibase 10.1103/PhysRevD.94.084005} {\bibfield  {journal} {\bibinfo
  {journal} {Phys. Rev.}\ }\textbf {\bibinfo {volume} {D94}},\ \bibinfo {pages}
  {084005} (\bibinfo {year} {2016})},\ \Eprint
  {http://arxiv.org/abs/1607.08460} {arXiv:1607.08460 [hep-th]} \BibitemShut
  {NoStop}%
\bibitem [{\citenamefont {Gonzalez-Martin}\ \emph {et~al.}(2017)\citenamefont
  {Gonzalez-Martin}, \citenamefont {Morris},\ and\ \citenamefont
  {Slade}}]{Gonzalez-Martin:2017gza}%
  \BibitemOpen
  \bibfield  {author} {\bibinfo {author} {\bibfnamefont {S.}~\bibnamefont
  {Gonzalez-Martin}}, \bibinfo {author} {\bibfnamefont {T.~R.}\ \bibnamefont
  {Morris}}, \ and\ \bibinfo {author} {\bibfnamefont {Z.~H.}\ \bibnamefont
  {Slade}},\ }\href {\doibase 10.1103/PhysRevD.95.106010} {\bibfield  {journal}
  {\bibinfo  {journal} {Phys. Rev.}\ }\textbf {\bibinfo {volume} {D95}},\
  \bibinfo {pages} {106010} (\bibinfo {year} {2017})},\ \Eprint
  {http://arxiv.org/abs/1704.08873} {arXiv:1704.08873 [hep-th]} \BibitemShut
  {NoStop}%
\bibitem [{\citenamefont {De~Brito}\ \emph {et~al.}(2018)\citenamefont
  {De~Brito}, \citenamefont {Ohta}, \citenamefont {Pereira}, \citenamefont
  {Tomaz},\ and\ \citenamefont {Yamada}}]{deBrito:2018jxt}%
  \BibitemOpen
  \bibfield  {author} {\bibinfo {author} {\bibfnamefont {G.~P.}\ \bibnamefont
  {De~Brito}}, \bibinfo {author} {\bibfnamefont {N.}~\bibnamefont {Ohta}},
  \bibinfo {author} {\bibfnamefont {A.~D.}\ \bibnamefont {Pereira}}, \bibinfo
  {author} {\bibfnamefont {A.~A.}\ \bibnamefont {Tomaz}}, \ and\ \bibinfo
  {author} {\bibfnamefont {M.}~\bibnamefont {Yamada}},\ }\href {\doibase
  10.1103/PhysRevD.98.026027} {\bibfield  {journal} {\bibinfo  {journal} {Phys.
  Rev.}\ }\textbf {\bibinfo {volume} {D98}},\ \bibinfo {pages} {026027}
  (\bibinfo {year} {2018})},\ \Eprint {http://arxiv.org/abs/1805.09656}
  {arXiv:1805.09656 [hep-th]} \BibitemShut {NoStop}%
\bibitem [{\citenamefont {Alkofer}(2019)}]{Alkofer:2018baq}%
  \BibitemOpen
  \bibfield  {author} {\bibinfo {author} {\bibfnamefont {N.}~\bibnamefont
  {Alkofer}},\ }\href {\doibase 10.1016/j.physletb.2018.12.061} {\bibfield
  {journal} {\bibinfo  {journal} {Phys. Lett.}\ }\textbf {\bibinfo {volume}
  {B789}},\ \bibinfo {pages} {480} (\bibinfo {year} {2019})},\ \Eprint
  {http://arxiv.org/abs/1809.06162} {arXiv:1809.06162 [hep-th]} \BibitemShut
  {NoStop}%
\bibitem [{\citenamefont {Falls}\ \emph {et~al.}(2019)\citenamefont {Falls},
  \citenamefont {Litim},\ and\ \citenamefont {Schröder}}]{Falls:2018ylp}%
  \BibitemOpen
  \bibfield  {author} {\bibinfo {author} {\bibfnamefont {K.~G.}\ \bibnamefont
  {Falls}}, \bibinfo {author} {\bibfnamefont {D.~F.}\ \bibnamefont {Litim}}, \
  and\ \bibinfo {author} {\bibfnamefont {J.}~\bibnamefont {Schröder}},\ }\href
  {\doibase 10.1103/PhysRevD.99.126015} {\bibfield  {journal} {\bibinfo
  {journal} {Phys. Rev.}\ }\textbf {\bibinfo {volume} {D99}},\ \bibinfo {pages}
  {126015} (\bibinfo {year} {2019})},\ \Eprint
  {http://arxiv.org/abs/1810.08550} {arXiv:1810.08550 [gr-qc]} \BibitemShut
  {NoStop}%
\bibitem [{\citenamefont {Don{\`a}}\ \emph {et~al.}(2014)\citenamefont
  {Don{\`a}}, \citenamefont {Eichhorn},\ and\ \citenamefont
  {Percacci}}]{Dona:2013qba}%
  \BibitemOpen
  \bibfield  {author} {\bibinfo {author} {\bibfnamefont {P.}~\bibnamefont
  {Don{\`a}}}, \bibinfo {author} {\bibfnamefont {A.}~\bibnamefont {Eichhorn}},
  \ and\ \bibinfo {author} {\bibfnamefont {R.}~\bibnamefont {Percacci}},\
  }\href {\doibase 10.1103/PhysRevD.89.084035} {\bibfield  {journal} {\bibinfo
  {journal} {Phys.Rev.}\ }\textbf {\bibinfo {volume} {D89}},\ \bibinfo {pages}
  {084035} (\bibinfo {year} {2014})},\ \Eprint {http://arxiv.org/abs/1311.2898}
  {arXiv:1311.2898 [hep-th]} \BibitemShut {NoStop}%
\bibitem [{\citenamefont {Meibohm}\ \emph {et~al.}(2016)\citenamefont
  {Meibohm}, \citenamefont {Pawlowski},\ and\ \citenamefont
  {Reichert}}]{Meibohm:2015twa}%
  \BibitemOpen
  \bibfield  {author} {\bibinfo {author} {\bibfnamefont {J.}~\bibnamefont
  {Meibohm}}, \bibinfo {author} {\bibfnamefont {J.~M.}\ \bibnamefont
  {Pawlowski}}, \ and\ \bibinfo {author} {\bibfnamefont {M.}~\bibnamefont
  {Reichert}},\ }\href {\doibase 10.1103/PhysRevD.93.084035} {\bibfield
  {journal} {\bibinfo  {journal} {Phys. Rev.}\ }\textbf {\bibinfo {volume}
  {D93}},\ \bibinfo {pages} {084035} (\bibinfo {year} {2016})},\ \Eprint
  {http://arxiv.org/abs/1510.07018} {arXiv:1510.07018 [hep-th]} \BibitemShut
  {NoStop}%
\bibitem [{\citenamefont {Don{\`a}}\ \emph {et~al.}(2016)\citenamefont
  {Don{\`a}}, \citenamefont {Eichhorn}, \citenamefont {Labus},\ and\
  \citenamefont {Percacci}}]{Dona:2015tnf}%
  \BibitemOpen
  \bibfield  {author} {\bibinfo {author} {\bibfnamefont {P.}~\bibnamefont
  {Don{\`a}}}, \bibinfo {author} {\bibfnamefont {A.}~\bibnamefont {Eichhorn}},
  \bibinfo {author} {\bibfnamefont {P.}~\bibnamefont {Labus}}, \ and\ \bibinfo
  {author} {\bibfnamefont {R.}~\bibnamefont {Percacci}},\ }\href {\doibase
  10.1103/PhysRevD.93.129904;;;;;;;;;;;;;;;;;;;;;;;;;;;;;;;;;;;;;;;
  10.1103/PhysRevD.93.044049} {\bibfield  {journal} {\bibinfo  {journal} {Phys.
  Rev.}\ }\textbf {\bibinfo {volume} {D93}},\ \bibinfo {pages} {044049}
  (\bibinfo {year} {2016})},\ \bibinfo {note} {[Erratum: Phys.
  Rev.D93,no.12,129904(2016)]},\ \Eprint {http://arxiv.org/abs/1512.01589}
  {arXiv:1512.01589 [gr-qc]} \BibitemShut {NoStop}%
\bibitem [{\citenamefont {Eichhorn}\ \emph
  {et~al.}(2018{\natexlab{a}})\citenamefont {Eichhorn}, \citenamefont {Labus},
  \citenamefont {Pawlowski},\ and\ \citenamefont
  {Reichert}}]{Eichhorn:2018akn}%
  \BibitemOpen
  \bibfield  {author} {\bibinfo {author} {\bibfnamefont {A.}~\bibnamefont
  {Eichhorn}}, \bibinfo {author} {\bibfnamefont {P.}~\bibnamefont {Labus}},
  \bibinfo {author} {\bibfnamefont {J.~M.}\ \bibnamefont {Pawlowski}}, \ and\
  \bibinfo {author} {\bibfnamefont {M.}~\bibnamefont {Reichert}},\ }\href
  {\doibase 10.21468/SciPostPhys.5.4.031} {\bibfield  {journal} {\bibinfo
  {journal} {SciPost Phys.}\ }\textbf {\bibinfo {volume} {5}},\ \bibinfo
  {pages} {31} (\bibinfo {year} {2018}{\natexlab{a}})},\ \Eprint
  {http://arxiv.org/abs/1804.00012} {arXiv:1804.00012 [hep-th]} \BibitemShut
  {NoStop}%
\bibitem [{\citenamefont {Alkofer}\ and\ \citenamefont
  {Saueressig}(2018)}]{Alkofer:2018fxj}%
  \BibitemOpen
  \bibfield  {author} {\bibinfo {author} {\bibfnamefont {N.}~\bibnamefont
  {Alkofer}}\ and\ \bibinfo {author} {\bibfnamefont {F.}~\bibnamefont
  {Saueressig}},\ }\href {\doibase 10.1016/j.aop.2018.07.017} {\bibfield
  {journal} {\bibinfo  {journal} {Annals Phys.}\ }\textbf {\bibinfo {volume}
  {396}},\ \bibinfo {pages} {173} (\bibinfo {year} {2018})},\ \Eprint
  {http://arxiv.org/abs/1802.00498} {arXiv:1802.00498 [hep-th]} \BibitemShut
  {NoStop}%
\bibitem [{\citenamefont {Henz}\ \emph {et~al.}(2013)\citenamefont {Henz},
  \citenamefont {Pawlowski}, \citenamefont {Rodigast},\ and\ \citenamefont
  {Wetterich}}]{Henz:2013oxa}%
  \BibitemOpen
  \bibfield  {author} {\bibinfo {author} {\bibfnamefont {T.}~\bibnamefont
  {Henz}}, \bibinfo {author} {\bibfnamefont {J.~M.}\ \bibnamefont {Pawlowski}},
  \bibinfo {author} {\bibfnamefont {A.}~\bibnamefont {Rodigast}}, \ and\
  \bibinfo {author} {\bibfnamefont {C.}~\bibnamefont {Wetterich}},\ }\href
  {\doibase 10.1016/j.physletb.2013.10.015} {\bibfield  {journal} {\bibinfo
  {journal} {Phys. Lett.}\ }\textbf {\bibinfo {volume} {B727}},\ \bibinfo
  {pages} {298} (\bibinfo {year} {2013})},\ \Eprint
  {http://arxiv.org/abs/1304.7743} {arXiv:1304.7743 [hep-th]} \BibitemShut
  {NoStop}%
\bibitem [{\citenamefont {Percacci}\ and\ \citenamefont
  {Vacca}(2015)}]{Percacci:2015wwa}%
  \BibitemOpen
  \bibfield  {author} {\bibinfo {author} {\bibfnamefont {R.}~\bibnamefont
  {Percacci}}\ and\ \bibinfo {author} {\bibfnamefont {G.~P.}\ \bibnamefont
  {Vacca}},\ }\href {\doibase 10.1140/epjc/s10052-015-3410-0} {\bibfield
  {journal} {\bibinfo  {journal} {Eur. Phys. J.}\ }\textbf {\bibinfo {volume}
  {C75}},\ \bibinfo {pages} {188} (\bibinfo {year} {2015})},\ \Eprint
  {http://arxiv.org/abs/1501.00888} {arXiv:1501.00888 [hep-th]} \BibitemShut
  {NoStop}%
\bibitem [{\citenamefont {Labus}\ \emph
  {et~al.}(2016{\natexlab{a}})\citenamefont {Labus}, \citenamefont {Percacci},\
  and\ \citenamefont {Vacca}}]{Labus:2015ska}%
  \BibitemOpen
  \bibfield  {author} {\bibinfo {author} {\bibfnamefont {P.}~\bibnamefont
  {Labus}}, \bibinfo {author} {\bibfnamefont {R.}~\bibnamefont {Percacci}}, \
  and\ \bibinfo {author} {\bibfnamefont {G.~P.}\ \bibnamefont {Vacca}},\ }\href
  {\doibase 10.1016/j.physletb.2015.12.022} {\bibfield  {journal} {\bibinfo
  {journal} {Phys. Lett.}\ }\textbf {\bibinfo {volume} {B753}},\ \bibinfo
  {pages} {274} (\bibinfo {year} {2016}{\natexlab{a}})},\ \Eprint
  {http://arxiv.org/abs/1505.05393} {arXiv:1505.05393 [hep-th]} \BibitemShut
  {NoStop}%
\bibitem [{\citenamefont {Oda}\ and\ \citenamefont
  {Yamada}(2016)}]{Oda:2015sma}%
  \BibitemOpen
  \bibfield  {author} {\bibinfo {author} {\bibfnamefont {K.-y.}\ \bibnamefont
  {Oda}}\ and\ \bibinfo {author} {\bibfnamefont {M.}~\bibnamefont {Yamada}},\
  }\href {\doibase 10.1088/0264-9381/33/12/125011} {\bibfield  {journal}
  {\bibinfo  {journal} {Class. Quant. Grav.}\ }\textbf {\bibinfo {volume}
  {33}},\ \bibinfo {pages} {125011} (\bibinfo {year} {2016})},\ \Eprint
  {http://arxiv.org/abs/1510.03734} {arXiv:1510.03734 [hep-th]} \BibitemShut
  {NoStop}%
\bibitem [{\citenamefont {Henz}\ \emph {et~al.}(2017)\citenamefont {Henz},
  \citenamefont {Pawlowski},\ and\ \citenamefont {Wetterich}}]{Henz:2016aoh}%
  \BibitemOpen
  \bibfield  {author} {\bibinfo {author} {\bibfnamefont {T.}~\bibnamefont
  {Henz}}, \bibinfo {author} {\bibfnamefont {J.~M.}\ \bibnamefont {Pawlowski}},
  \ and\ \bibinfo {author} {\bibfnamefont {C.}~\bibnamefont {Wetterich}},\
  }\href {\doibase 10.1016/j.physletb.2017.01.057} {\bibfield  {journal}
  {\bibinfo  {journal} {Phys. Lett.}\ }\textbf {\bibinfo {volume} {B769}},\
  \bibinfo {pages} {105} (\bibinfo {year} {2017})},\ \Eprint
  {http://arxiv.org/abs/1605.01858} {arXiv:1605.01858 [hep-th]} \BibitemShut
  {NoStop}%
\bibitem [{\citenamefont {Wetterich}\ and\ \citenamefont
  {Yamada}(2017)}]{Wetterich:2016uxm}%
  \BibitemOpen
  \bibfield  {author} {\bibinfo {author} {\bibfnamefont {C.}~\bibnamefont
  {Wetterich}}\ and\ \bibinfo {author} {\bibfnamefont {M.}~\bibnamefont
  {Yamada}},\ }\href {\doibase 10.1016/j.physletb.2017.04.049} {\bibfield
  {journal} {\bibinfo  {journal} {Phys. Lett.}\ }\textbf {\bibinfo {volume}
  {B770}},\ \bibinfo {pages} {268} (\bibinfo {year} {2017})},\ \Eprint
  {http://arxiv.org/abs/1612.03069} {arXiv:1612.03069 [hep-th]} \BibitemShut
  {NoStop}%
\bibitem [{\citenamefont {Biemans}\ \emph {et~al.}(2017)\citenamefont
  {Biemans}, \citenamefont {Platania},\ and\ \citenamefont
  {Saueressig}}]{Biemans:2017zca}%
  \BibitemOpen
  \bibfield  {author} {\bibinfo {author} {\bibfnamefont {J.}~\bibnamefont
  {Biemans}}, \bibinfo {author} {\bibfnamefont {A.}~\bibnamefont {Platania}}, \
  and\ \bibinfo {author} {\bibfnamefont {F.}~\bibnamefont {Saueressig}},\
  }\href {\doibase 10.1007/JHEP05(2017)093} {\bibfield  {journal} {\bibinfo
  {journal} {JHEP}\ }\textbf {\bibinfo {volume} {05}},\ \bibinfo {pages} {093}
  (\bibinfo {year} {2017})},\ \Eprint {http://arxiv.org/abs/1702.06539}
  {arXiv:1702.06539 [hep-th]} \BibitemShut {NoStop}%
\bibitem [{\citenamefont {Hamada}\ and\ \citenamefont
  {Yamada}(2017)}]{Hamada:2017rvn}%
  \BibitemOpen
  \bibfield  {author} {\bibinfo {author} {\bibfnamefont {Y.}~\bibnamefont
  {Hamada}}\ and\ \bibinfo {author} {\bibfnamefont {M.}~\bibnamefont
  {Yamada}},\ }\href {\doibase 10.1007/JHEP08(2017)070} {\bibfield  {journal}
  {\bibinfo  {journal} {JHEP}\ }\textbf {\bibinfo {volume} {08}},\ \bibinfo
  {pages} {070} (\bibinfo {year} {2017})},\ \Eprint
  {http://arxiv.org/abs/1703.09033} {arXiv:1703.09033 [hep-th]} \BibitemShut
  {NoStop}%
\bibitem [{\citenamefont {Becker}\ \emph {et~al.}(2017)\citenamefont {Becker},
  \citenamefont {Ripken},\ and\ \citenamefont {Saueressig}}]{Becker:2017tcx}%
  \BibitemOpen
  \bibfield  {author} {\bibinfo {author} {\bibfnamefont {D.}~\bibnamefont
  {Becker}}, \bibinfo {author} {\bibfnamefont {C.}~\bibnamefont {Ripken}}, \
  and\ \bibinfo {author} {\bibfnamefont {F.}~\bibnamefont {Saueressig}},\
  }\href {\doibase 10.1007/JHEP12(2017)121} {\bibfield  {journal} {\bibinfo
  {journal} {JHEP}\ }\textbf {\bibinfo {volume} {12}},\ \bibinfo {pages} {121}
  (\bibinfo {year} {2017})},\ \Eprint {http://arxiv.org/abs/1709.09098}
  {arXiv:1709.09098 [hep-th]} \BibitemShut {NoStop}%
\bibitem [{\citenamefont {Eichhorn}\ \emph
  {et~al.}(2018{\natexlab{b}})\citenamefont {Eichhorn}, \citenamefont
  {Lippoldt},\ and\ \citenamefont {Skrinjar}}]{Eichhorn:2017sok}%
  \BibitemOpen
  \bibfield  {author} {\bibinfo {author} {\bibfnamefont {A.}~\bibnamefont
  {Eichhorn}}, \bibinfo {author} {\bibfnamefont {S.}~\bibnamefont {Lippoldt}},
  \ and\ \bibinfo {author} {\bibfnamefont {V.}~\bibnamefont {Skrinjar}},\
  }\href {\doibase 10.1103/PhysRevD.97.026002} {\bibfield  {journal} {\bibinfo
  {journal} {Phys. Rev.}\ }\textbf {\bibinfo {volume} {D97}},\ \bibinfo {pages}
  {026002} (\bibinfo {year} {2018}{\natexlab{b}})},\ \Eprint
  {http://arxiv.org/abs/1710.03005} {arXiv:1710.03005 [hep-th]} \BibitemShut
  {NoStop}%
\bibitem [{\citenamefont {Eichhorn}\ \emph
  {et~al.}(2018{\natexlab{c}})\citenamefont {Eichhorn}, \citenamefont {Hamada},
  \citenamefont {Lumma},\ and\ \citenamefont {Yamada}}]{Eichhorn:2017als}%
  \BibitemOpen
  \bibfield  {author} {\bibinfo {author} {\bibfnamefont {A.}~\bibnamefont
  {Eichhorn}}, \bibinfo {author} {\bibfnamefont {Y.}~\bibnamefont {Hamada}},
  \bibinfo {author} {\bibfnamefont {J.}~\bibnamefont {Lumma}}, \ and\ \bibinfo
  {author} {\bibfnamefont {M.}~\bibnamefont {Yamada}},\ }\href {\doibase
  10.1103/PhysRevD.97.086004} {\bibfield  {journal} {\bibinfo  {journal} {Phys.
  Rev.}\ }\textbf {\bibinfo {volume} {D97}},\ \bibinfo {pages} {086004}
  (\bibinfo {year} {2018}{\natexlab{c}})},\ \Eprint
  {http://arxiv.org/abs/1712.00319} {arXiv:1712.00319 [hep-th]} \BibitemShut
  {NoStop}%
\bibitem [{\citenamefont {Pawlowski}\ \emph {et~al.}(2019)\citenamefont
  {Pawlowski}, \citenamefont {Reichert}, \citenamefont {Wetterich},\ and\
  \citenamefont {Yamada}}]{Pawlowski:2018ixd}%
  \BibitemOpen
  \bibfield  {author} {\bibinfo {author} {\bibfnamefont {J.~M.}\ \bibnamefont
  {Pawlowski}}, \bibinfo {author} {\bibfnamefont {M.}~\bibnamefont {Reichert}},
  \bibinfo {author} {\bibfnamefont {C.}~\bibnamefont {Wetterich}}, \ and\
  \bibinfo {author} {\bibfnamefont {M.}~\bibnamefont {Yamada}},\ }\href
  {\doibase 10.1103/PhysRevD.99.086010} {\bibfield  {journal} {\bibinfo
  {journal} {Phys. Rev.}\ }\textbf {\bibinfo {volume} {D99}},\ \bibinfo {pages}
  {086010} (\bibinfo {year} {2019})},\ \Eprint
  {http://arxiv.org/abs/1811.11706} {arXiv:1811.11706 [hep-th]} \BibitemShut
  {NoStop}%
\bibitem [{\citenamefont {Knorr}\ \emph {et~al.}(2019)\citenamefont {Knorr},
  \citenamefont {Ripken},\ and\ \citenamefont {Saueressig}}]{Knorr:2019atm}%
  \BibitemOpen
  \bibfield  {author} {\bibinfo {author} {\bibfnamefont {B.}~\bibnamefont
  {Knorr}}, \bibinfo {author} {\bibfnamefont {C.}~\bibnamefont {Ripken}}, \
  and\ \bibinfo {author} {\bibfnamefont {F.}~\bibnamefont {Saueressig}},\
  }\href {\doibase 10.1088/1361-6382/ab4a53} {\bibfield  {journal} {\bibinfo
  {journal} {Class. Quant. Grav.}\ }\textbf {\bibinfo {volume} {36}},\ \bibinfo
  {pages} {234001} (\bibinfo {year} {2019})},\ \Eprint
  {http://arxiv.org/abs/1907.02903} {arXiv:1907.02903 [hep-th]} \BibitemShut
  {NoStop}%
\bibitem [{\citenamefont {Wetterich}(2019{\natexlab{b}})}]{Wetterich:2019rsn}%
  \BibitemOpen
  \bibfield  {author} {\bibinfo {author} {\bibfnamefont {C.}~\bibnamefont
  {Wetterich}},\ }\href@noop {} {\  (\bibinfo {year} {2019}{\natexlab{b}})},\
  \Eprint {http://arxiv.org/abs/1911.06100} {arXiv:1911.06100 [hep-th]}
  \BibitemShut {NoStop}%
\bibitem [{\citenamefont {Christiansen}\ \emph
  {et~al.}(2018{\natexlab{b}})\citenamefont {Christiansen}, \citenamefont
  {Litim}, \citenamefont {Pawlowski},\ and\ \citenamefont
  {Reichert}}]{Christiansen:2017cxa}%
  \BibitemOpen
  \bibfield  {author} {\bibinfo {author} {\bibfnamefont {N.}~\bibnamefont
  {Christiansen}}, \bibinfo {author} {\bibfnamefont {D.~F.}\ \bibnamefont
  {Litim}}, \bibinfo {author} {\bibfnamefont {J.~M.}\ \bibnamefont
  {Pawlowski}}, \ and\ \bibinfo {author} {\bibfnamefont {M.}~\bibnamefont
  {Reichert}},\ }\href {\doibase 10.1103/PhysRevD.97.106012} {\bibfield
  {journal} {\bibinfo  {journal} {Phys. Rev.}\ }\textbf {\bibinfo {volume}
  {D97}},\ \bibinfo {pages} {106012} (\bibinfo {year} {2018}{\natexlab{b}})},\
  \Eprint {http://arxiv.org/abs/1710.04669} {arXiv:1710.04669 [hep-th]}
  \BibitemShut {NoStop}%
\bibitem [{\citenamefont {Litim}\ and\ \citenamefont
  {Pawlowski}(1998)}]{Litim:1998qi}%
  \BibitemOpen
  \bibfield  {author} {\bibinfo {author} {\bibfnamefont {D.~F.}\ \bibnamefont
  {Litim}}\ and\ \bibinfo {author} {\bibfnamefont {J.~M.}\ \bibnamefont
  {Pawlowski}},\ }\href {\doibase 10.1016/S0370-2693(98)00761-8} {\bibfield
  {journal} {\bibinfo  {journal} {Phys.Lett.}\ }\textbf {\bibinfo {volume}
  {B435}},\ \bibinfo {pages} {181} (\bibinfo {year} {1998})},\ \Eprint
  {http://arxiv.org/abs/hep-th/9802064} {arXiv:hep-th/9802064 [hep-th]}
  \BibitemShut {NoStop}%
\bibitem [{\citenamefont {Ellwanger}(1994)}]{Ellwanger:1993mw}%
  \BibitemOpen
  \bibfield  {author} {\bibinfo {author} {\bibfnamefont {U.}~\bibnamefont
  {Ellwanger}},\ }\bibfield  {booktitle} {\emph {\bibinfo {booktitle}
  {{Proceedings, Workshop on Quantum field theoretical aspects of high energy
  physics: Bad Frankenhausen, Germany, September 20-24, 1993}}},\ }\href
  {\doibase 10.1007/BF01555911} {\bibfield  {journal} {\bibinfo  {journal} {Z.
  Phys.}\ }\textbf {\bibinfo {volume} {C62}},\ \bibinfo {pages} {503} (\bibinfo
  {year} {1994})},\ \Eprint {http://arxiv.org/abs/hep-ph/9308260}
  {arXiv:hep-ph/9308260 [hep-ph]} \BibitemShut {NoStop}%
\bibitem [{\citenamefont {Morris}(1994)}]{Morris:1993qb}%
  \BibitemOpen
  \bibfield  {author} {\bibinfo {author} {\bibfnamefont {T.~R.}\ \bibnamefont
  {Morris}},\ }\href {\doibase 10.1142/S0217751X94000972} {\bibfield  {journal}
  {\bibinfo  {journal} {Int. J. Mod. Phys.}\ }\textbf {\bibinfo {volume}
  {A9}},\ \bibinfo {pages} {2411} (\bibinfo {year} {1994})},\ \Eprint
  {http://arxiv.org/abs/hep-ph/9308265} {arXiv:hep-ph/9308265} \BibitemShut
  {NoStop}%
\bibitem [{\citenamefont {Rubin}\ and\ \citenamefont
  {Ordonez}(1985)}]{Rubin:1984tc}%
  \BibitemOpen
  \bibfield  {author} {\bibinfo {author} {\bibfnamefont {M.~A.}\ \bibnamefont
  {Rubin}}\ and\ \bibinfo {author} {\bibfnamefont {C.~R.}\ \bibnamefont
  {Ordonez}},\ }\href {\doibase 10.1063/1.526749} {\bibfield  {journal}
  {\bibinfo  {journal} {J. Math. Phys.}\ }\textbf {\bibinfo {volume} {26}},\
  \bibinfo {pages} {65} (\bibinfo {year} {1985})}\BibitemShut {NoStop}%
\bibitem [{\citenamefont {Camporesi}\ and\ \citenamefont
  {Higuchi}(1994)}]{Camporesi:1994ga}%
  \BibitemOpen
  \bibfield  {author} {\bibinfo {author} {\bibfnamefont {R.}~\bibnamefont
  {Camporesi}}\ and\ \bibinfo {author} {\bibfnamefont {A.}~\bibnamefont
  {Higuchi}},\ }\href {\doibase 10.1063/1.530850} {\bibfield  {journal}
  {\bibinfo  {journal} {J. Math. Phys.}\ }\textbf {\bibinfo {volume} {35}},\
  \bibinfo {pages} {4217} (\bibinfo {year} {1994})}\BibitemShut {NoStop}%
\bibitem [{\citenamefont {Lippoldt}(2018)}]{Lippoldt:2018wvi}%
  \BibitemOpen
  \bibfield  {author} {\bibinfo {author} {\bibfnamefont {S.}~\bibnamefont
  {Lippoldt}},\ }\href {\doibase 10.1016/j.physletb.2018.05.037} {\bibfield
  {journal} {\bibinfo  {journal} {Phys. Lett.}\ }\textbf {\bibinfo {volume}
  {B782}},\ \bibinfo {pages} {275} (\bibinfo {year} {2018})},\ \Eprint
  {http://arxiv.org/abs/1804.04409} {arXiv:1804.04409 [hep-th]} \BibitemShut
  {NoStop}%
\bibitem [{\citenamefont {Falls}\ \emph
  {et~al.}(2018{\natexlab{b}})\citenamefont {Falls}, \citenamefont {King},
  \citenamefont {Litim}, \citenamefont {Nikolakopoulos},\ and\ \citenamefont
  {Rahmede}}]{Falls:2017lst}%
  \BibitemOpen
  \bibfield  {author} {\bibinfo {author} {\bibfnamefont {K.}~\bibnamefont
  {Falls}}, \bibinfo {author} {\bibfnamefont {C.~R.}\ \bibnamefont {King}},
  \bibinfo {author} {\bibfnamefont {D.~F.}\ \bibnamefont {Litim}}, \bibinfo
  {author} {\bibfnamefont {K.}~\bibnamefont {Nikolakopoulos}}, \ and\ \bibinfo
  {author} {\bibfnamefont {C.}~\bibnamefont {Rahmede}},\ }\href {\doibase
  10.1103/PhysRevD.97.086006} {\bibfield  {journal} {\bibinfo  {journal} {Phys.
  Rev.}\ }\textbf {\bibinfo {volume} {D97}},\ \bibinfo {pages} {086006}
  (\bibinfo {year} {2018}{\natexlab{b}})},\ \Eprint
  {http://arxiv.org/abs/1801.00162} {arXiv:1801.00162 [hep-th]} \BibitemShut
  {NoStop}%
\bibitem [{\citenamefont {Litim}\ and\ \citenamefont
  {Pawlowski}(2002{\natexlab{a}})}]{Litim:2002ce}%
  \BibitemOpen
  \bibfield  {author} {\bibinfo {author} {\bibfnamefont {D.~F.}\ \bibnamefont
  {Litim}}\ and\ \bibinfo {author} {\bibfnamefont {J.~M.}\ \bibnamefont
  {Pawlowski}},\ }\href@noop {} {\bibfield  {journal} {\bibinfo  {journal}
  {JHEP}\ }\textbf {\bibinfo {volume} {0209}},\ \bibinfo {pages} {049}
  (\bibinfo {year} {2002}{\natexlab{a}})},\ \Eprint
  {http://arxiv.org/abs/hep-th/0203005} {arXiv:hep-th/0203005 [hep-th]}
  \BibitemShut {NoStop}%
\bibitem [{\citenamefont {Litim}\ and\ \citenamefont
  {Pawlowski}(2002{\natexlab{b}})}]{Litim:2002hj}%
  \BibitemOpen
  \bibfield  {author} {\bibinfo {author} {\bibfnamefont {D.~F.}\ \bibnamefont
  {Litim}}\ and\ \bibinfo {author} {\bibfnamefont {J.~M.}\ \bibnamefont
  {Pawlowski}},\ }\href {\doibase 10.1016/S0370-2693(02)02693-X} {\bibfield
  {journal} {\bibinfo  {journal} {Phys.Lett.}\ }\textbf {\bibinfo {volume}
  {B546}},\ \bibinfo {pages} {279} (\bibinfo {year} {2002}{\natexlab{b}})},\
  \Eprint {http://arxiv.org/abs/hep-th/0208216} {arXiv:hep-th/0208216 [hep-th]}
  \BibitemShut {NoStop}%
\bibitem [{\citenamefont {Pawlowski}(2007)}]{Pawlowski:2005xe}%
  \BibitemOpen
  \bibfield  {author} {\bibinfo {author} {\bibfnamefont {J.~M.}\ \bibnamefont
  {Pawlowski}},\ }\href {\doibase 10.1016/j.aop.2007.01.007} {\bibfield
  {journal} {\bibinfo  {journal} {Annals Phys.}\ }\textbf {\bibinfo {volume}
  {322}},\ \bibinfo {pages} {2831} (\bibinfo {year} {2007})},\ \Eprint
  {http://arxiv.org/abs/hep-th/0512261} {arXiv:hep-th/0512261 [hep-th]}
  \BibitemShut {NoStop}%
\bibitem [{\citenamefont {Folkerts}\ \emph {et~al.}(2012)\citenamefont
  {Folkerts}, \citenamefont {Litim},\ and\ \citenamefont
  {Pawlowski}}]{Folkerts:2011jz}%
  \BibitemOpen
  \bibfield  {author} {\bibinfo {author} {\bibfnamefont {S.}~\bibnamefont
  {Folkerts}}, \bibinfo {author} {\bibfnamefont {D.~F.}\ \bibnamefont {Litim}},
  \ and\ \bibinfo {author} {\bibfnamefont {J.~M.}\ \bibnamefont {Pawlowski}},\
  }\href {\doibase 10.1016/j.physletb.2012.02.002} {\bibfield  {journal}
  {\bibinfo  {journal} {Phys.Lett.}\ }\textbf {\bibinfo {volume} {B709}},\
  \bibinfo {pages} {234} (\bibinfo {year} {2012})},\ \Eprint
  {http://arxiv.org/abs/1101.5552} {arXiv:1101.5552 [hep-th]} \BibitemShut
  {NoStop}%
\bibitem [{\citenamefont {Donkin}\ and\ \citenamefont
  {Pawlowski}(2012)}]{Donkin:2012ud}%
  \BibitemOpen
  \bibfield  {author} {\bibinfo {author} {\bibfnamefont {I.}~\bibnamefont
  {Donkin}}\ and\ \bibinfo {author} {\bibfnamefont {J.~M.}\ \bibnamefont
  {Pawlowski}},\ }\href@noop {} {\  (\bibinfo {year} {2012})},\ \Eprint
  {http://arxiv.org/abs/1203.4207} {arXiv:1203.4207 [hep-th]} \BibitemShut
  {NoStop}%
\bibitem [{\citenamefont {Bridle}\ \emph {et~al.}(2014)\citenamefont {Bridle},
  \citenamefont {Dietz},\ and\ \citenamefont {Morris}}]{Bridle:2013sra}%
  \BibitemOpen
  \bibfield  {author} {\bibinfo {author} {\bibfnamefont {I.~H.}\ \bibnamefont
  {Bridle}}, \bibinfo {author} {\bibfnamefont {J.~A.}\ \bibnamefont {Dietz}}, \
  and\ \bibinfo {author} {\bibfnamefont {T.~R.}\ \bibnamefont {Morris}},\
  }\href {\doibase 10.1007/JHEP03(2014)093} {\bibfield  {journal} {\bibinfo
  {journal} {JHEP}\ }\textbf {\bibinfo {volume} {03}},\ \bibinfo {pages} {093}
  (\bibinfo {year} {2014})},\ \Eprint {http://arxiv.org/abs/1312.2846}
  {arXiv:1312.2846 [hep-th]} \BibitemShut {NoStop}%
\bibitem [{\citenamefont {Dietz}\ and\ \citenamefont
  {Morris}(2015)}]{Dietz:2015owa}%
  \BibitemOpen
  \bibfield  {author} {\bibinfo {author} {\bibfnamefont {J.~A.}\ \bibnamefont
  {Dietz}}\ and\ \bibinfo {author} {\bibfnamefont {T.~R.}\ \bibnamefont
  {Morris}},\ }\href {\doibase 10.1007/JHEP04(2015)118} {\bibfield  {journal}
  {\bibinfo  {journal} {JHEP}\ }\textbf {\bibinfo {volume} {04}},\ \bibinfo
  {pages} {118} (\bibinfo {year} {2015})},\ \Eprint
  {http://arxiv.org/abs/1502.07396} {arXiv:1502.07396 [hep-th]} \BibitemShut
  {NoStop}%
\bibitem [{\citenamefont {Safari}(2016)}]{Safari:2015dva}%
  \BibitemOpen
  \bibfield  {author} {\bibinfo {author} {\bibfnamefont {M.}~\bibnamefont
  {Safari}},\ }\href {\doibase 10.1140/epjc/s10052-016-4036-6} {\bibfield
  {journal} {\bibinfo  {journal} {Eur. Phys. J.}\ }\textbf {\bibinfo {volume}
  {C76}},\ \bibinfo {pages} {201} (\bibinfo {year} {2016})},\ \Eprint
  {http://arxiv.org/abs/1508.06244} {arXiv:1508.06244 [hep-th]} \BibitemShut
  {NoStop}%
\bibitem [{\citenamefont {Labus}\ \emph
  {et~al.}(2016{\natexlab{b}})\citenamefont {Labus}, \citenamefont {Morris},\
  and\ \citenamefont {Slade}}]{Labus:2016lkh}%
  \BibitemOpen
  \bibfield  {author} {\bibinfo {author} {\bibfnamefont {P.}~\bibnamefont
  {Labus}}, \bibinfo {author} {\bibfnamefont {T.~R.}\ \bibnamefont {Morris}}, \
  and\ \bibinfo {author} {\bibfnamefont {Z.~H.}\ \bibnamefont {Slade}},\ }\href
  {\doibase 10.1103/PhysRevD.94.024007} {\bibfield  {journal} {\bibinfo
  {journal} {Phys. Rev.}\ }\textbf {\bibinfo {volume} {D94}},\ \bibinfo {pages}
  {024007} (\bibinfo {year} {2016}{\natexlab{b}})},\ \Eprint
  {http://arxiv.org/abs/1603.04772} {arXiv:1603.04772 [hep-th]} \BibitemShut
  {NoStop}%
\bibitem [{\citenamefont {Eichhorn}\ and\ \citenamefont
  {Held}(2018{\natexlab{a}})}]{Eichhorn:2017ylw}%
  \BibitemOpen
  \bibfield  {author} {\bibinfo {author} {\bibfnamefont {A.}~\bibnamefont
  {Eichhorn}}\ and\ \bibinfo {author} {\bibfnamefont {A.}~\bibnamefont
  {Held}},\ }\href {\doibase 10.1016/j.physletb.2017.12.040} {\bibfield
  {journal} {\bibinfo  {journal} {Phys. Lett.}\ }\textbf {\bibinfo {volume}
  {B777}},\ \bibinfo {pages} {217} (\bibinfo {year} {2018}{\natexlab{a}})},\
  \Eprint {http://arxiv.org/abs/1707.01107} {arXiv:1707.01107 [hep-th]}
  \BibitemShut {NoStop}%
\bibitem [{\citenamefont {Eichhorn}\ and\ \citenamefont
  {Held}(2018{\natexlab{b}})}]{Eichhorn:2018whv}%
  \BibitemOpen
  \bibfield  {author} {\bibinfo {author} {\bibfnamefont {A.}~\bibnamefont
  {Eichhorn}}\ and\ \bibinfo {author} {\bibfnamefont {A.}~\bibnamefont
  {Held}},\ }\href {\doibase 10.1103/PhysRevLett.121.151302} {\bibfield
  {journal} {\bibinfo  {journal} {Phys. Rev. Lett.}\ }\textbf {\bibinfo
  {volume} {121}},\ \bibinfo {pages} {151302} (\bibinfo {year}
  {2018}{\natexlab{b}})},\ \Eprint {http://arxiv.org/abs/1803.04027}
  {arXiv:1803.04027 [hep-th]} \BibitemShut {NoStop}%
\bibitem [{\citenamefont {Reichert}\ and\ \citenamefont
  {Smirnov}(2019)}]{Reichert:2019car}%
  \BibitemOpen
  \bibfield  {author} {\bibinfo {author} {\bibfnamefont {M.}~\bibnamefont
  {Reichert}}\ and\ \bibinfo {author} {\bibfnamefont {J.}~\bibnamefont
  {Smirnov}},\ }\href@noop {} {\  (\bibinfo {year} {2019})},\ \Eprint
  {http://arxiv.org/abs/1911.00012} {arXiv:1911.00012 [hep-ph]} \BibitemShut
  {NoStop}%
\end{thebibliography}%
\end{document}